\newcommand\vx{\vec{x}}

\newcommand\hr{\hat{r}}

\newcommand\rs{r_{\rm s}}
\newcommand\deltam{\delta_{\rm m}}

\newcommand\deltagal{\delta_{\rm g}}

\newcommand\omegam{\Omega_{\rm m}}
\newcommand\omegab{\Omega_{\rm b}}

\newcommand\oO{\mathcal{O}}

\newcommand\Vmax{V_{\rm max}}
\newcommand\Rmax{R_{\rm max}}
\newcommand{\specialcell}[2][l]{%
  \begin{tabular}[#1]{@{}l@{}}#2\end{tabular}}

\newcommand\DV{D_{\rm V}}

\newcommand\Mpch{{\rm\; Mpc}/h}

\newcommand{\tF}{\tilde{F}_2}

\documentclass[useAMS,usenatbib]{mn2e}

\usepackage{placeins}

\usepackage{graphicx} 
\usepackage{multirow}
\usepackage{array}

\usepackage{amsmath,amssymb}

\usepackage[T1]{fontenc}
\usepackage[latin9]{inputenc}
\usepackage{float}
\usepackage{graphicx}
\usepackage{esint}

\makeatletter
\DeclareFontEncoding{LGR}{}{}
\DeclareTextSymbol{\~}{LGR}{126}

\begin{document}

\title[Baryon Acoustic Oscillations in the 3PCF]{Detection of Baryon Acoustic Oscillation Features in the Large-Scale 3-Point Correlation Function of SDSS BOSS DR12 CMASS Galaxies}

\author{\makeatauthor}
\author[Slepian et al.]{Zachary Slepian$^{1}$\thanks{zslepian@cfa.harvard.edu},
Daniel J. Eisenstein$^{1}$\thanks{deisenstein@cfa.harvard.edu},
Joel R. Brownstein$^{2}$,
Chia-Hsun Chuang$^{3,4}$, \and
H\'ector Gil-Mar\'in$^{5, 6}$,
Shirley Ho$^{7,8,9}$,
Francisco-Shu Kitaura$^4$,
Will J. Percival$^{10}$,\and
Ashley J. Ross$^{11}$,
Graziano Rossi$^{12}$,
Hee-Jong Seo$^{13}$,
An\v{z}e Slosar$^{14}$\and
\& Mariana Vargas-Maga\~na$^{15}$\\
$^{1}$ Harvard-Smithsonian Center for Astrophysics, 60 Garden Street, Cambridge, MA 02138, USA\\
$^{2}$Department of Physics and Astronomy, University of Utah, Salt Lake City, UT 84112, USA\\
$^3$ Instituto de F\'{\i}sica Te\'orica, (UAM/CSIC), Universidad Aut\'onoma de Madrid, Cantoblanco, E-28049 Madrid, Spain \\
$^4$ Leibniz-Institut f\"{u}r Astrophysik Potsdam (AIP), An der Sternwarte 16, D-14482 Potsdam, Germany\\
$^{5}$ Sorbonne Universit\'es, Institut Lagrange de Paris (ILP), 98 bis Boulevard Arago, 75014 Paris, France\\
$^{6}$ Laboratoire de Physique Nucl\'eaire et de Hautes Energies, Universit\'e Pierre et Marie Curie, 4 Place Jussieu, 75005 Paris, France\\
$^7$ Lawrence Berkeley National Lab, 1 Cyclotron Rd, Berkeley CA 94720, USA\\
$^8$ McWilliams Center for Cosmology, Department of Physics, Carnegie Mellon University, 5000 Forbes Ave., Pittsburgh, PA 15213, USA\\
$^9$ Department of Physics, University of California, Berkeley, CA 94720, USA\\
$^{10}$ Institute of Cosmology \& Gravitation, University of Portsmouth, Dennis Sciama Building, Portsmouth PO1 3FX, UK\\
$^{11}$ Center for Cosmology and Astroparticle Physics, Department of Physics, The Ohio State University, OH 43210, USA\\
$^{12}$ Department of Physics and Astronomy, Sejong University, Seoul, 143-747, Korea\\
$^{13}$ Department of Physics and Astronomy, Ohio University, Clippinger Labs, Athens, OH 45701, USA\\
$^{14}$ Brookhaven National Laboratory, Upton, NY 11973, USA\\
$^{15}$ Instituto de F\'isica, Universidad Nacional Aut\'onoma de M\'exico, Apdo. Postal 20-364, M\'exico}

\maketitle

\begin{abstract}
We present the large-scale 3-point correlation function (3PCF) of the SDSS DR12 CMASS sample of $777,202$ Luminous Red Galaxies, the largest-ever sample used for a 3PCF or bispectrum measurement. We make the first high-significance ($4.5\sigma$) detection of Baryon Acoustic Oscillations (BAO) in the 3PCF. Using these acoustic features in the 3PCF as a standard ruler, we measure the distance to $z=0.57$ to $1.7\%$ precision (statistical plus systematic).  We find $\DV = 2024\pm29\;{\rm Mpc\;(stat)}\pm20\;{\rm Mpc\;(sys)}$ for our fiducial cosmology (consistent with {\it Planck} 2015) and bias model. This measurement extends the use of the BAO technique from the 2-point correlation function (2PCF) and power spectrum to the 3PCF and opens an avenue for deriving additional cosmological distance information from future large-scale structure redshift surveys such as DESI. Our measured distance scale from the 3PCF is fairly independent from that derived from the pre-reconstruction 2PCF and is equivalent to increasing the length of BOSS by roughly 10\%; reconstruction appears to lower the independence of the distance measurements.  Fitting a model including tidal tensor bias yields a moderate significance ($2.6\sigma)$ detection of this bias with a value in agreement with the prediction from local Lagrangian biasing.

\end{abstract}

\section{Introduction}
\label{sec:intro}
Determining the nature of dark energy is one of the most pressing problems of modern cosmology.  Efforts have focused on measuring the dark energy equation of state $w$, which is $-1$ if dark energy is a cosmological constant; any other value of $w$ means the dark energy density evolves in time (Copeland, Sami \& Tsujikawa 2006).  The dark energy density dictates the Universe's expansion through the Friedmann equation, and so measuring the Universe's size as a function of time or redshift constrains the equation of state. A number of techniques exist to do this (Weinberg et al. 2013), one of the most prominent being the Baryon Acoustic Oscillation (BAO) method.

The BAO method exploits a preferred scale imprinted on the baryon density at decoupling ($z\sim1020$). Prior to decoupling, the Universe is a hot, dense, ionized plasma in which electrons couple to photons by Thomson scattering and protons follow electrons under the Coulomb force.  Primordially overdense regions are overpressured, and the radiation pressure, dominant at high redshift, launches spherical pressure-density (sound) waves of baryons and photons outwards from each overdensity at roughly $c/\sqrt{3}$ (Sakharov 1966; Peebles \& Yu 1970; Sunyaev \& Zel'dovich 1970; Bond \& Efstathiou 1984, 1987; Holtzmann 1989; Hu \& Sugiyama 1996; Eisenstein \& Hu 1998; Eisenstein, Seo \& White 2007).  These are the BAO, and the sound waves travel outwards until decoupling, where they halt as the photons precipitously release them since Thomson scattering is no longer effective. At the wavefront the baryon velocity is maximal and the late-time growing mode inherits the spatial structure of the velocity. The BAO thus correlate the original overdensity with a sharp excess density of baryons a sound horizon (roughly $\rs \approx 100\Mpch$ comoving) away.  

Once the Universe is neutral, on large scales the baryons and the dark matter experience only gravity, so the two components converge and the excess density of baryons imprints on the total matter density (Hu \& Sugiyama 1996; Eisenstein \& Hu 1998; Slepian \& Eisenstein 2016a).  When galaxies begin to form, they trace the matter density field and so the BAO produce a slight excess of galaxy pairs separated by $\sim 100\Mpch$. This excess translates to a sharp, localized BAO bump in the 2-point correlation function (2PCF) of galaxies, which measures the excess probability over random of finding one galaxy at a given separation from another; there are analogous BAO features in the 2PCF's Fourier-space analog the power spectrum.

Since the BAO signal is produced by large-scale, pre-decoupling physics, it is frozen into the comoving distribution of galaxies. Consequently measuring the BAO scale from galaxy clustering in different redshift slices provides a differential history of the Universe's expansion rate (Eisenstein, Hu \& Tegmark 1998; Blake \& Glazebrook 2003; Hu \& Haiman 2003; Linder 2003; Seo \& Eisenstein 2003). The BAO scale is also imprinted on the temperature anisotropies in the Cosmic Microwave Background (CMB), since the density structure at that epoch determined the temperature. The CMB therefore offers an absolute scale for the BAO method.

Thus far, the BAO method has used the 2PCF of galaxies as well as the galaxy power spectrum to measure the cosmic distance scale to high precision.  Since the original detections of the BAO bump in the 2PCF of galaxies (Cole et al. 2005; Eisenstein et al. 2005), large-scale redshift surveys such as the Sloan Digital Sky Survey (SDSS) and Baryon Oscillation Spectroscopic Survey (BOSS) have yielded ever-increasing precision via the BAO method. The current precision on the distance scale from the 2PCF/power spectrum is of order $1\%$ (Anderson et al. 2014; Cuesta et al. 2016; Gil-Mar\'in et al. 2016), and future surveys such as Dark Energy Spectroscopic Instrument (DESI) (Levi et al. 2013) and {\it Euclid} (Laureijs et al. 2011) should achieve a factor of five improvement in precision. The Lyman-$\alpha$ forest has also been used for BAO measurements, with the most recent results in Delubac et al. (2015). The first detection of BAO in voids has also recently been made, offering an additional possible avenue to the distance scale (Kitaura et al. 2016).

Until now the BAO method has not explicitly used higher correlations of the galaxy density field.  As earlier noted, the BAO produce an excess of pairs of galaxies separated by $100\Mpch$, but the BAO also imprint on triplets of galaxies, creating a slight excess of triangles where one or more triangle side is of the BAO scale. Triplets develop correlations both due to non-linear structure formation and non-linear bias. Slepian \& Eisenstein (2016b; hereafter SE16b) shows that there are distinctive BAO features in the 3-point correlation function (3PCF) of galaxies.  Detecting these features would enable a measurement of the cosmic distance scale from the 3PCF alone.  

Thus far, only two previous works have measured the 3PCF on physical scales large enough to access the BAO scale.  Gazta\~{n}aga et al. (2009) used a sample of $\sim 40,000$ Luminous Red Galaxies (LRGs) from SDSS DR7. They find a $2-3\sigma$ detection of the BAO using all opening angles of a single triangle configuration with side lengths $r_1 = 33 \Mpch$ and $r_2 = 88\Mpch$. Slepian et al. (2015; hereafter S15) used $777,202$ LRGs from the CMASS sample within SDSS-III BOSS to measure the 3PCF in a compressed basis where many triangle configurations were used but one of the two sides was integrated out over a wedge set by the remaining free side.  That work found a $2.8\sigma$ detection of the BAO.  Given the larger sample of S15, by comparison to Gazta\~naga et al. (2009) a higher significance BAO detection might be expected, suggesting that there is BAO information the compressed basis does not exploit.  On the other hand, Gazta\~naga et al. (2009) did find an anomalously high baryon fraction (roughly double the presently-accepted value), which would increase the significance of a BAO detection.  Neither work used these moderate-significance BAO detections to measure the cosmic distance scale.

In this work, we use the same dataset and 3PCF measurement as S15. However, we do not compress by integrating out one triangle side.  The compression scheme of S15 was motivated by avoiding any triangle side's becoming small and two galaxies becoming close, where linear perturbation theory is likely a poor model. Here, we avoid this limit by choosing triangle sides such that the smallest side never is below $20\Mpch$.  

We again use the novel algorithm of Slepian \& Eisenstein (2015b,c; hereafter SE15b,c), which computes the 3PCF's multipole moments in $\oO(Nn\Vmax)$ time using spherical harmonic decompositions, where $\Vmax$ is the volume of a sphere of radius $\Rmax$, the maximum triangle side length to which correlations are measured. The covariance matrix also turns out to be tractable in the multipole basis (SE15b). The main outcomes of this work are:

\indent{\bf 1)} The first high-significance $(\sim 4.5\sigma)$ detection of the BAO in the 3PCF.\\
\indent {\bf 2)} A measurement of the cosmic distance scale at redshift $0.57$ to $1.7\%$ precision from the 3PCF.\\
\indent {\bf 3)} High precision ($\sim 1\%$) determination of the linear bias at fixed $\sigma_8$ for this sample from the 3PCF.\\

An interesting subsidiary result of this work is that the tidal tensor bias $b_t$ (further detailed in \S\ref{sec:models}) of the dataset agrees well with the theoretically predicted relation with linear bias $b_1$, $b_t = -(2/7)[b_1-1]$ (Baldauf et al. 2012; Chan, Scoccimarro \& Sheth 2012), offering mild evidence for the validity of local Lagrangian biasing. In contrast, $b_t$ for the {\textsc PATCHY} mock catalogs for SDSS DR12 (described further in \S\ref{sec:data_randoms_mocks}) does not agree with this theoretical relation. With our work's error bars on $b_t$, the tension between mocks and data is only mild, but this possible misfit between the data and the {\textsc PATCHY} mocks as well as the {\textsc PATCHY} mocks and the theory may warrant further investigation.

The paper is laid out as follows.  \S\ref{sec:data_randoms_mocks} details our dataset, the random catalogs used for edge correction, and the mock catalogs used to obtain parameters within the covariance matrix as well as to verify our pipeline.  \S\ref{sec:method} summarizes the multipole basis we use for the 3PCF as well as the algorithm of SE15b used for the measurement, while \S\ref{sec:covar} discusses our covariance matrix.  In \S\ref{sec:models} we outline the two different bias models we use to analyze the data, a ``minimal'' model that includes linear and non-linear biasing, and a ``tidal tensor'' model that includes these elements and also tidal tensor biasing. \S\ref{sec:fitting_procedure} details our parameter-fitting procedure, and \S\ref{sec:BAO_fit} presents our BAO detection and best-fit parameters for the data and mocks.  \S\ref{sec:distance_scale} gives our distance scale measurement in physical units and compares with other recent works, while \S\ref{sec:bias_parameters} discusses our measured bias parameters. We conclude in \S\ref{sec:concs}.

\section{Data, Randoms, and Mocks}
\label{sec:data_randoms_mocks}
Here we introduce the dataset, random catalogs, and mock catalogs used for this work as well as giving details on the SDSS and BOSS.  We used the CMASS sample (Alam et al. 2015) within SDSS BOSS DR12 (Eisenstein et al. 2011; Dawson et al. 2013), comprising  $777,202$ LRGs.  CMASS denotes that the sample was color-selected to have roughly constant stellar mass, with $M_* >10^{11}\;M_{\odot}$. The survey totals 9,493 square degrees (Reid et al. 2016), with roughly $73\%$ of the area and galaxies in the North Galactic Cap and the remainder in the South Galactic Cap; the redshift range is $0.43$ to $0.7$.  Further details of the target selection and catalog construction are given in Reid et al. (2016), with observational systematic biases discussed in Ross et al. (2012; 2015).  

Overall, the SDSS (York et al. 2000) was divided into three parts: SDSS I and II (Abazajian et al. 2009) and SDSS III (Eisenstein et al. 2011).  Using a drift-scanning mosaic CCD camera (Gunn et al. 1998) on the 2.5-m Sloan Telescope (Gunn et al. 2006) at Apache Point Observatory in New Mexico, the survey imaged $14,555$ square degrees in five photometric bandpasses (Fukugita et al. 1996; Smith et al. 2002; Doi et al. 2010).  Details of the astrometric calibration are given in Pier et al. (2003); the photometric reduction in Lupton et al. (2001), and the photometric calibration in Padmanabhan et al. (2008). The entire dataset was reprocessed for Data Release 8 as described in Aihara et al. (2011).  For BOSS specifically, target assignment was performed using an adaptive algorithm outlined in Blanton et al. (2003) and spectroscopy via double-armed spectrographs (Smee et al. 2013). Redshifts were then derived as detailed in Bolton et al. (2012).

To verify our covariance matrix, as well as to test our analysis pipeline and assess the typicality of our results from the data, we also computed the 3PCF of 298 mock catalogs developed for DR12 known as the \textsc{MultiDark-Patchy} BOSS DR12 mocks (Kitaura, Yepes \& Prada 2014; Kitaura et al. 2015a,b). We passed these mocks through the same pipeline as the data including fitting bias models to them and considering the BAO significance and distance information. 

Briefly, these catalogs used second-order Lagrangian perturbation theory (2LPT) combined with a spherical collapse model on small scales (Kitaura \& He{\ss} 2013), and were calibrated on accurate N-body-based reference catalogs. The calibration used halo abundance matching to reproduce the number density, clustering bias, selection function, and survey geometry of the BOSS data (Rodr\'iguez-Torres et al. 2015).

\section{Method}
\label{sec:method}
This work uses the basis of Legendre polynomials $P_{\ell}$ for the dependence of the 3PCF on triangle opening angle; we thus measure the 3PCF at each multipole as a function of the two triangle sides (Szapudi 2004; SE15a). Mathematically the full 3PCF $\zeta$ is expanded as
\begin{align}
\zeta(r_1, r_2;\hr_1\cdot\hr_2) = \sum_{\ell}\zeta_{\ell}(r_1, r_2)P_{\ell}(\hr_1\cdot\hr_2),
\end{align}
where $P_{\ell}$ is a Legendre polynomial.

We use the novel 3PCF algorithm of SE15b,c to measure the multipole moments of the 3PCF.  For details of the multipole basis and algorithm we refer the reader to SE15b,c, with a shorter summary in S15; here we simply recapitulate its major advantages.

First, the multipole basis underlies our algorithm to measure the 3PCF in a way scaling as $Nn\Vmax\sim N^2$, where $N$ is the number of galaxies in the survey and $n$ is the number density. A naive triplet count would scale as $N(n\Vmax)^2\sim N^3$. There have been other 3PCF algorithms that improve upon the naive triple count, but they involve approximations and are not highly efficient for measuring the 3PCF on large scales. As shown in SE15b, the algorithm is 500 times faster than a triplet count and only 6 times slower than a 2PCF computation. It allows us to measure the 3PCF for of order one million galaxies in a few minutes on modest computing resources (runtimes are further detailed in SE15b \S5).

The speed advantage occurs because our algorithm relies on the computation of angular cross-power spectra between spherical shells centered iteratively on each galaxy in the survey. It has long been known how to quickly estimate the angular auto power spectrum of a given shell around one origin in the context of CMB analyses. Given that these analyses use a large maximum multipole of order several thousand, for the CMB it is more efficient to grid the temperature anisotropy field and take a spherical harmonic transform scaling as $G^{3/2}$, with $G$ the number of grid cells.  The spherical harmonic coefficients can then be assembled into the multipole moments (i.e. angular power spectrum). Since for the 3PCF we only require a modest number of multipoles, our algorithm simply computes the spherical harmonic coefficients directly, which is more accurate since the density field never needs to be gridded.

Second, the multipole basis permits analytic calculation of the covariance matrix in terms of simple 2-D integrals of the linear power spectrum and spherical Bessel functions (SE15b \S6) if one assumes the dominant contribution is from an underlying Gaussian-random field plus shot noise and ignores all connected terms in the 6-point function.  This covariance matrix avoids the noise and consequent non-invertibility of a covariance matrix determined solely from
mocks.  To use this latter strategy one would need to determine $\ell_{\rm max}^2N_{\rm bins}^2/2$ independent matrix elements; for the present work this number is $20,000$.  Furthermore, it is desirable to have many more than one mock per dimension of the covariance matrix to avoid noise (Percival et al. 2014). Determining an invertible covariance matrix from mocks alone would thus require computing the 3PCF of a large number of mocks. Consequently, in previous works on the 3PCF or bispectrum, a variety of assumptions about the structure of the covariance matrix and its eigenvalues have been adopted (e.g. Gazta\~naga et al. 2009; Gil-Mar\'in et al. 2015).  Having an analytic covariance matrix avoids need for approximations in this regard.  

Third, edge-correction is straightforward in the multipole basis. Details are presented in SE15b \S4. Here we  note that measuring the 3PCF of a catalog of a large number of random points thrown in the survey volume is sufficient for accurate edge correction. The edge correction can be cast as a matrix inversion and performed as a post-processing step that takes negligible time.

\section{Covariance}
\label{sec:covar}
We use an analytic covariance matrix as computed in SE15b \S6, with volume $V$ and survey number density $n$ (shot noise scales as $1/n$) as free parameters to be fit from an empirical covariance matrix derived from $298$ mock catalogs.\footnote{In our previous work on the CMASS 3PCF, S15, we used 299 mocks, but further analysis revealed that one mock ($\#132$) was corrupted and we do not use it here.}  As earlier noted, our covariance matrix calculation assumes a Gaussian random field for the density, but this assumption is reasonable given that the density field on large scales is only very weakly non-Gaussian even at low redshifts.  

The covariance calculation is succinctly summarized in S15 \S6, as is our procedure for fitting the best survey volume and number density from the mock catalogs. This latter procedure uses a likelihood metric proposed by Xu et al. (2012).  

To test our analytic covariance matrix $\bf{C}_{\rm GRF}$, we construct its half-inverse $\bf{C}_{\rm GRF}^{-1/2}$ and apply symmetrically to the mock covariance matrix $\bf{C}_{\rm mock}$.  If our analytic covariance matrix accurately describes the true independence structure of the 3PCF of the mocks, then we should have $\bf{C}_{\rm GRF}^{-1/2}C_{\rm mock}\bf{C}_{\rm GRF}^{-1/2} - \bf{I} =0$, where $\bf{I}$ is the identity matrix. A major advantage of this test is that it avoids ever inverting the mocks' covariance matrix, which as discussed in \S\ref{sec:method} will be noisy at best and non-invertible at worst.

\begin{figure}
\centering
\includegraphics[width=.54\textwidth]{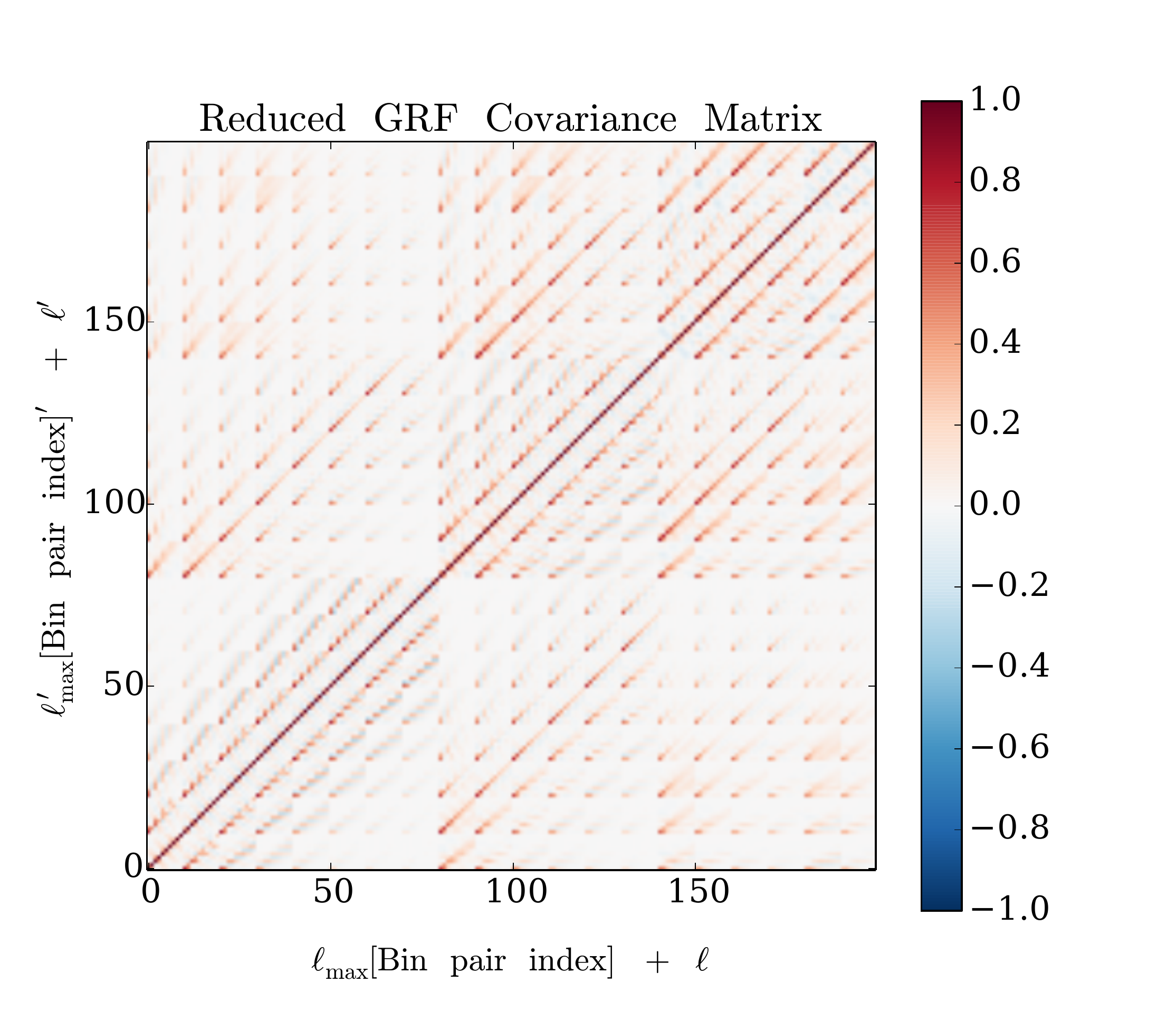}
\caption{The reduced covariance matrix computed as outlined in \S\ref{sec:covar}, with volume and shot noise fixed by fitting to the mocks' empirical covariance matrix (though note the volume divides out of the reduced covariance). The reduced covariance matrix is the full covariance matrix divided by the geometric mean of the diagonal elements: $C_{{\rm red,}\;ij} \equiv C_{ij}/\sqrt{C_{ii}C_{jj}}$. Here multipole varies faster than bin pair; thus the small tiles visible in the matrix are all multipoles for a given bin pair.}
\label{fig:reduced_covar}
\end{figure}

\begin{figure}
\centering
\includegraphics[width=.54\textwidth]{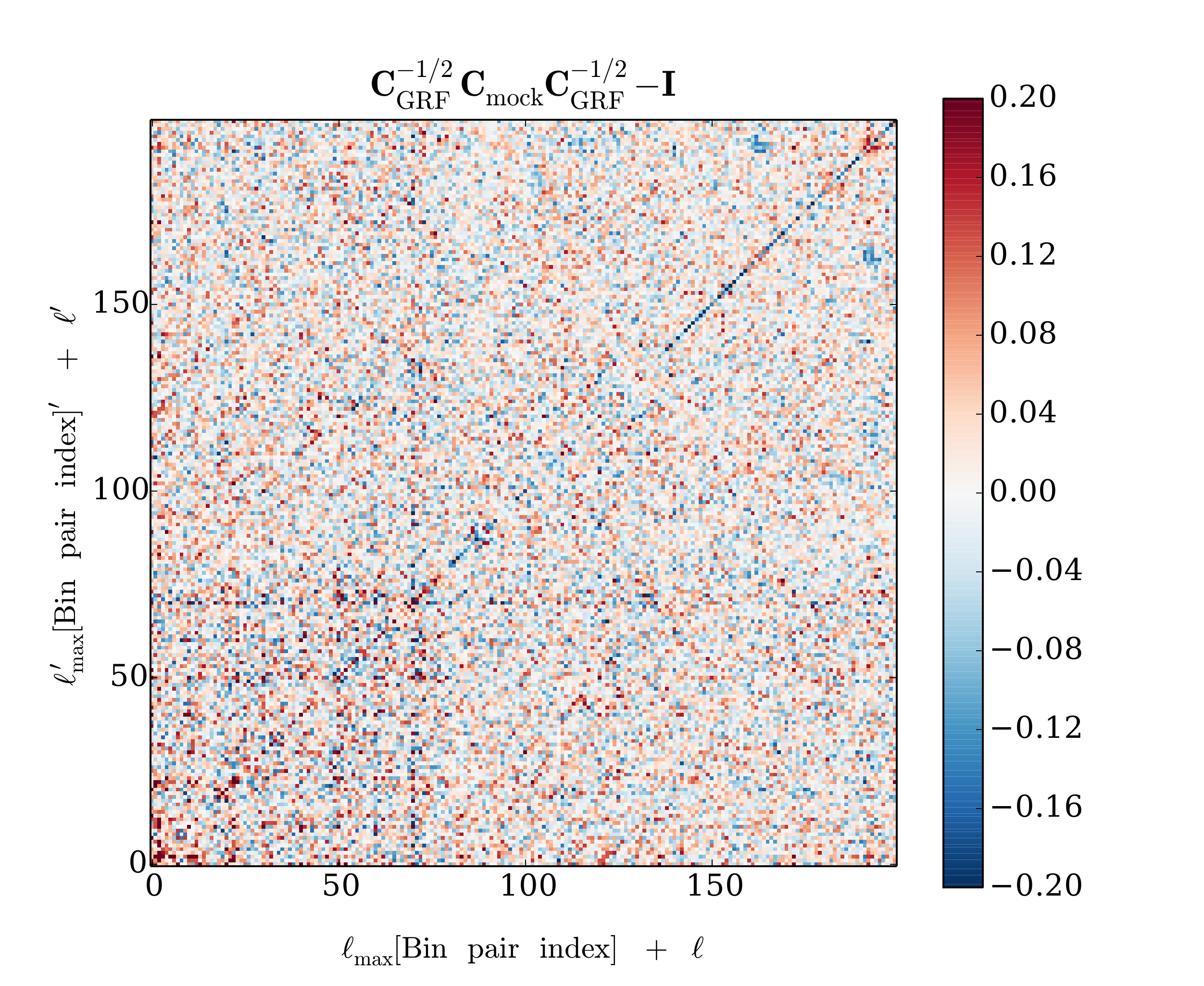}
\caption{Our test of the analytic Gaussian random field covariance against an empirical covariance matrix estimated from 298 mocks. We have subtracted the identity matrix; were the test perfect the mean would be zero. In reality the mean is $0.004$. Given the 298 mocks, we expect root mean square noise in this plot of $0.058$, as discussed in \S\ref{sec:covar}; in reality it is $0.062$.}
\label{fig:covar_test}
\end{figure}

In Figure \ref{fig:reduced_covar}, we show our analytic reduced covariance matrix with the best-fit volume $V=2.43\;{\rm [Gpc}/h]^3$ and number density $n=1.41\times 10^{-4}\;{\rm [Mpc/}h]^{-3}$. In Figure \ref{fig:covar_test}, we show the half-inverse test; it should look reasonably close to random noise. Given that we used a finite number of mocks to estimate the matrix elements of $\bf{C}_{\rm mock}$, even were our analytic covariance matrix a perfect match to the true covariance of the mocks' 3PCF, we would expect a root mean square scatter of $1/\sqrt{298}\approx 0.058$.  As an additional test we do not plot, we examined the eigenvalues of the mocks' covariance matrix and the analytic covariance matrix and found fairly good agreement; the deviations were consistent with noise from the small number of mocks used. We also computed the average ratio of the diagonal of the analytic covariance matrix to the diagonal of the empirical covariance formed from the mocks, finding $0.97$, which for $200$ elements is within $1\sigma$ of unity, where $\sigma \simeq 1/\sqrt{200} = 0.07$.

\section{Modeling the 3PCF}
\label{sec:models}
\subsection{Minimal and tidal tensor bias models}
\label{subsec:bias_models}
We fit two different models to the data in this work. First, we consider a model where the galaxy overdensity field $\deltagal$ traces the matter density field $\deltam$ and its square with two unknown bias coefficients, the linear bias $b_1$ and the non-linear bias $b_2$. This bias model is
\begin{align}
\deltagal(\vx) = b_1\deltam(\vx) + b_2\left[\deltam^2(\vx) - \left<\deltam^2(\vx) \right>\right],
\end{align}
where the expectation value is over translations and must be subtracted so that $\left< \deltagal \right>=0$.  To obtain the leading order (fourth order) 3PCF, the matter density field must be expanded to second order as
\begin{align}
\deltam(\vx) = \delta(\vx) + \delta^{(2)}(\vx).
\end{align}
$\delta$ is the linear density field and $\delta^{(2)}$ is $\oO(\delta^2)$ and computed by integrating two copies of the Fourier-transformed linear density field at different wavenumbers against the perturbation theory kernel $\tF$ (Bernardeau et al. 2002).  This kernel encodes the evolution of the density under Newtonian gravity. For Gaussian-random-field initial conditions, this evolution is the reason the late-time field is non-Gaussian on large scales and has a non-vanishing large-scale 3PCF.  

Second, we consider a model where the galaxy overdensity traces the matter density field, its square, and the local tidal tensor $s(\vx)$ with three unknown bias coefficients, $b_1$, $b_2$, and $b_t$ (Fry 1996; Catelan et al. 1998; Catelan et al. 2000; McDonald \& Roy 2009; Baldauf et al. 2012; Chan, Scoccimarro \& Sheth 2012). This bias model is
\begin{align}
\deltagal(\vx) = & \;   b_1\deltam(\vx) + b_2\left[\deltam^2(\vx) - \left<\deltam^2(\vx) \right>\right] \nonumber\\
&  + b_t\left[ s^2(\vx) - \left<s^2(\vx) \right>\right],
\end{align}
where again we subtract expectation values where necessary to ensure $\left<\deltagal \right>=0$. The local Lagrangian bias model finds that $b_t = -(2/7)[b_1-1]$, though this is only valid for large-scale modes and does not incorporate renormalization (McDonald 2006).


One can compute the 3PCF in two steps. First, one chooses a particular galaxy to sit at the origin. To simplify the calculation one then takes it that this galaxy always contributes the most complicated bias term to the triple product $\left<\deltagal(\vx_1)\deltagal(\vx_2)\deltagal(\vx_3) \right>$. The result of this computation is the pre-cyclic 3PCF. In reality, we do not know which galaxy sits at the origin of coordinates, or equivalently, contributes a particular bias term. Consequently, the second step in computing the 3PCF is   cyclically summing over all three possible choices of origin.  This yields the full, post-cyclic 3PCF to be compared to observations. For further discussion, see SE15a and SE16b.

In the multipole basis, the real-space pre-cyclic 3PCF is remarkably compact: it has only monopole $(\ell=0)$, dipole $(\ell=1)$, and quadrupole $(\ell=2)$ moments, as discussed in SE15a,b and references therein.  In redshift space, even at the pre-cyclic level all multipoles enter, but those at $\ell\geq 3$ are all $\oO(\beta^2)$ at leading order (SE16b), where $\beta = f/b_1\sim 0.4$ for the CMASS sample, with $f= d\ln D/(d\ln a)\approx \omegam^{0.55}$, $D$ the linear growth rate and $a$ the scale factor. Thus even in redshift space most of the pre-cyclic structure is set by the monopole, dipole, and quadrupole, as further discussed in SE16b. Regarding redshift space, we also note that the model of SE16b is fully self-consistent to leading order in the 3PCF (i.e. fourth order in the linear density field) and is not merely a Kaiser treatment of RSD, as is the bispectrum model of Scoccimarro, Couchman \& Frieman (1998) on which SE16b is based.

Cyclic summing then mixes each pre-cyclic multipole into all post-cyclic multipoles, as discussed in SE15a \S7 and SE16b \S4.1.  As regards the BAO, the post-cyclic $\ell=0$ receives most of its structure from the pre-cyclic $\ell=0$, with a sub-dominant contribution from the pre-cyclic $\ell=1$. The post-cyclic $\ell=1$ receives most of its structure from the pre-cyclic $\ell=1$, with a sub-dominant contribution from the pre-cyclic $\ell=0$.  The post-cyclic $\ell=2$ receives roughly equal contributions in amplitude from the pre-cyclic $\ell=1$ and $\ell=2$, but essentially all of the BAO structure in the post-cyclic $\ell=2$ comes from the pre-cyclic $\ell=1$.  For all of the higher post-cyclic multipoles, again the BAO structure comes almost entirely from the pre-cyclic $\ell=1$.  These couplings are shown in SE16b Figure 6.  

We note that the higher post-cyclic multipoles ($\ell \geq 3$) all look rather similar to each other, as shown in SE16b Figures 6-8; this is because the information is all coming from a compact set of pre-cyclic multipoles.  Consequently we believe there is not significant additional information in the higher multipoles. Thus while in principle the 3PCF has an infinite number of multipoles, in practice most of the information is likely in the first few. Thus the multipole basis is a relatively compact basis for measuring the 3PCF and we likely do not lose much by stopping at $\ell_{\rm max} = 9$ as we do in this work.

The full equations for the pre-cyclic model we use are given in SE16b equation (21); these are then cyclically summed and reprojected onto the multipoles as discussed above.

\subsection{Integral constraint}
\label{subsec:integral_const}

Finally, we model a failure of the integral constraint in our fitting.  The integral constraint demands that the average number density of randoms equal the average number density of galaxies, as discussed in more detail in S15 \S5.3.  We thus perturb our 3PCF estimator by rescaling the randoms by $1+c$, and find that this induces an additional term in the measured 3PCF $\zeta^{\rm M}$ as
\begin{align}
\zeta^{\rm M}=\frac{\zeta^{\rm T} -c\xi_{\rm cyc}-c^3}{(1+c)^3},
\end{align}
as shown in S15 equations (22) and (23); $\zeta^{\rm T}$ is the true 3PCF. $\xi_{\rm cyc}$ is the cyclic sum of the linear theory 2PCF $\xi$ around the three triangle sides, $\xi_{\rm cyc}  = \xi(r_1)+\xi(r_2)+\xi(r_3)$. Anomalous large-scale power near the survey scale, such as from wide-angle observational systematics, would produce a similar distortion. Though the rescaling of the randoms is as $1+c$, moving forward, we will track $c$ as the integral constraint amplitude.

\subsection{Power spectra}
\label{subsec:power_spectra}
As shown in SE16b, the full 3PCF model is simply a set of integral transforms of the power spectrum.  In the present work, we use two different template power spectra.  First, we use the physical power spectrum $P_{\rm phys}$ to compute the models. Second, we use the no-wiggle power spectrum, which has the large-scale CDM growth correct but with all BAO features removed. We use these two power spectrum templates because comparing the $\chi^2$ from fitting the data to each bias model with no-wiggle and physical power spectra gives a measure of the BAO significance in the data.

The physical power spectrum is the linear power spectrum with the BAO component smoothed by a Gaussian to model RSD and non-linear structure formation (Eisenstein, Seo \& White 2007; Anderson et al. 2012):
\begin{align}
P_{\rm phys}(k) = \left[P(k) - P_{\rm nw}(k) \right]\exp\left [ -k^2\Sigma_{\rm nl}^2/2\right] +P_{\rm nw}(k),
\end{align}
with $P$ the linear theory power spectrum. $P_{\rm nw}$ is the no-wiggle power spectrum and is computed from the fitting formula for the no-wiggle transfer function given in Eisenstein \& Hu (1998).  $\Sigma_{\rm nl}=8\Mpch$ is the non-linear smoothing scale.  

The power spectrum is always formed from the matter transfer function $T_{\rm m}$ as $P=Ak^{n_{\rm s}}T_{\rm m}^2$, where $A$ is an amplitude set by fixing $\sigma_8$ today. We compute the transfer functions from Code for Anisotropies in the Cosmic Microwave Background (\textsc{CAMB}; Lewis 2000) using a geometrically flat $\Lambda{\rm  CDM}$ cosmology with parameters matching those used for the \textsc{MultiDark-Patchy} mock catalogs (Kitaura et al. 2015) and consistent with the Planck values (Planck Paper XIII, 2015). Our cosmology also matches that used for S15. 

The parameters are $\omegab = 0.048,\;\omegam = 0.307115,\;h\equiv H_0/(100\;{\rm km/s/Mpc})=0.6777,\;n_{\rm s}=0.9611,\;\sigma_8(z=0)=0.8288,\;T_{\rm CMB} = 2.7255\;{\rm K}$. We take the survey redshift to be $z_{\rm survey} = 0.57$, the average of the range $0.43<z<0.7$; this is a good approximation because the redshift distribution of objects is roughly symmetric about the middle of this interval (Reid et al. 2015). We rescale $\sigma_8$ by the ratio of the linear growth factor at the survey redshift to the linear growth factor at redshift zero. 

\subsection{Varying $\alpha$}
A final aspect of our model is to vary the effective size of the Universe at our survey redshift.  Within both the physical and no-wiggle templates, we allow the physical distances to be rescaled by  a factor $\alpha$, where $\alpha = 1$ if our assumed fiducial cosmology is correct. Given a 3PCF at multipole $\ell$ for the fiducial cosmology $\zeta^{\rm F}_{\ell}(r_1, r_2)$, our model 3PCF with varying $\alpha$ will be  $\zeta^{\rm \mathcal{M}}_{\ell}(r_1,  r_2)= \zeta^{\rm F}_{\ell}(\alpha r_1, \alpha r_2)$.

The sense of our rescaling is that $\alpha < 1$ moves the BAO features to larger physical scales in the model, while $\alpha > 1$ moves the BAO features to smaller physical scales in the model. For instance, a best fit value of $\alpha<1$ will associate the observed BAO with a model where they appear on larger scales than the fiducial cosmology. For both the physical and no-wiggle templates, we use $\alpha$ in the range $0.8<\alpha<1.25$.  

Varying $\alpha$ allows us to measure the cosmic distance scale at the survey redshift relative to the sound horizon for the fiducial cosmology recorded above.  Our dilation or contraction rescales the 3PCF amplitude by roughly $(1-4\alpha)$, since $\zeta \sim 1/(r_1^2 r_2^2) +{\rm cyc.}$ and $r_1,r_2\to \alpha r_1,\alpha r_2$. Since we do not renormalize to a fixed $\sigma_8$ after dilation or contraction, changing $\alpha$ induces a shift in $b_1$, causing a substantial correlation of $b_1$ with $\alpha$.

\section{Fitting procedure}
\label{sec:fitting_procedure}
\subsection{Triangle configurations used}
We briefly outline the triangle configurations used and then turn to our high-significance BAO detection. As discussed in \S\ref{sec:intro} we wish to avoid triangles where any two galaxies are too close to each other such that non-linear structure formation has become important and linear perturbation theory likely provides a poor model.  To acheive this we use all bins in $r_1,\;r_2$ where the minimum of any side is $>20\Mpch$ and the maximum of any side is $<140\Mpch$. The $20\Mpch$ minimum is dictated by avoiding squeezed triangles; the maximum reflects a decision we have made that there is very limited signal to noise in larger scale bins. 

Furthermore, there may be as-yet unresolved large-scale systematics in the survey that become dominant on these scales. In particular, we computed the half-inverse covariance matrix test for a number of different maximal scales and found that the analytic covariance matrix did not reproduce that derived from the mocks as well on larger scales.  The choice of $140\Mpch$ as a maximal scale was thus dictated by the likelihood of diminishing returns from larger scales and the concern that the covariance matrix on larger scales was not as well-controlled. 

Explicitly, our criteria hold for the twenty bin combinations in the set ${\bf S}=$ \big\{[2, 5], [2, 6], [2, 7], [2, 8], [2, 9], [2, 10], [2, 11], [2, 12], [3, 6], [3, 7], [3, 8], [3, 9], [3, 10], [3, 11], [4, 7], [4, 8], [4, 9], [4, 10], [5, 8], [5, 9]\big\}. Bin 0 in $r_1$ would mean $0\leq r_1 < 10\Mpch$, bin 1 in $r_1$ would mean $10\leq r_1 < 20\Mpch$, etc., and analogously for $r_2$.
\subsection{Bias parameters, $\beta$, and integral constraint amplitude}
\label{subsec:bias_fitting}
We briefly describe our procedure for fitting the free parameters of the models presented in \S\ref{sec:models} to the data, as there are some dimensions of the problem that can be significantly accelerated due to the structure of the models.  In particular, at fixed $\alpha,\;\beta,$ and $c$, for the bias parameters ($b_1, \;b_2,$ and $b_t$), our model is a sum with terms proportional to $b_1^3, \; b_1^2b_2,$ and $b_1^2b_t$.   These three combinations are independent, and thus the minimum $\chi^2$ for the total model is the sum of the minimum $\chi^2$ for each of these three terms.  

Consequently rather than doing an expensive 3-D search in the space $(b_1,\; b_2,\; b_t)$ for our tidal tensor model or in the 2-D space $(b_1,\;b_2)$ for our minimal model, we can solve directly for the best $b_1$, $b_2$, and $b_t$ as a least-squares minimization Gaussian likelihood problem. For each $\alpha, \;\beta,$ and $c$, this procedure gives the best-fit biases. The procedure also returns the covariance matrix of these parameters ${\bf C}_{\rm biases}$ as a function of $\alpha,\;\beta$, and $c$. 

Unfortunately, $\alpha$, $\beta$, and $c$ do not enter our model linearly (see SE16b equation (21)). Therefore we require explicit loops over them.  We explored fitting for $\beta$ but found essentially no constraint; it is highly degenerate with $b_1$, as SE16b equation (21) suggests.  We therefore elected to set $\beta = f/b_1$ using $f\approx \omegam^{0.55}(z_{\rm survey} =0.57)$. Consequently in this work the linear bias determines the value of $\beta$, meaning that once we have recovered $b_1$ from our fit we must ensure that the $\beta$ used was consistent with it.  Since $b_1$ depends on both the model (minimal or tidal tensor) and the dataset (true data or mocks), we consider four values of $\beta$ in this work.  For the data, we use $\beta = 0.43$ for the minimal model and $\beta = 0.37$ for the tidal tensor model; for the mocks, we use $\beta = 0.40$ for the minimal model and $\beta = 0.49$ for the tidal tensor model. $\beta$ for the mocks is computed using the average values of $b_1$ over all mocks in each model. We suppress $\beta$ as an argument in \S\ref{subsec:bias_marginalization} and \S\ref{subsec:alpha_marginalization}. 

To fit the integral constraint amplitude $c$, we subtract our integral constraint model from the data itself and from the model. This subtraction casts the model fully in terms linear in the biases being fit, as detailed above. We use a grid of $31$ values of the integral constraint amplitude $c$ in the range $-0.03$ to $0.03$.

\subsection{Bias marginalization}
\label{subsec:bias_marginalization}

Our fitting results in a surface of $\chi^2$ vs. $\alpha$ and $c$, as well as the best-fit biases and their covariance matrices at all values of $\alpha$ and $c$. We fit the dataset as well as 298 mock catalogs; the runtime is a negligible fraction of the full 3PCF calculation.  

Given the $\chi^2$ surface, we wish to marginalize over the integral constraint amplitude $c$.  We also wish to include the Gaussian error bars on the biases in this marginalization, as these do depend slightly on the integral constraint amplitude.  We evaluate
\begin{align}
&\left<b_i^n(\alpha)\right> =\frac{1}{N_{b_i}(\alpha)} \nonumber\\
&\times \int dc\;db_i\;b_i^n\exp\left[-\frac{\chi^2(\alpha,c)}{2}-\frac{(b_i-b_{i,{\rm best}}(\alpha,c))^2}{2\sigma_{b_i}^2(\alpha,c)}\right]
\label{eqn:bias_expectation_value}
\end{align}
where $N_{b_i}$ is the normalization given by equation (\ref{eqn:bias_expectation_value}) evaluated with $n=0$.
$b_{i,{\rm best}}(\alpha,c)$ is the best-fit bias at a given $\alpha$ and $c$, while $\sigma_{b_i}^2(\alpha,c)$ is the square root of the appropriate diagonal element of the bias covariance matrix at that $\alpha$ and $c$. We have neglected a factor of $\det \bf{C}_{\rm biases}$ in these integrals; this factor would account for changes in the covariance matrix with $\alpha$ and $c$. However, the covariance matrix does not change rapidly enough with $\alpha$ and $c$ to argue for this complication. 

Setting $n=1$ in equation (\ref{eqn:bias_expectation_value}) gives the bias $b_i$ marginalized over $c$, and setting $n=2$ gives the expectation value of its squate. Using these two quantities we have the marginalized root mean square of the biases as
\begin{align}
\sigma(b_i(\alpha)) = \sqrt{\left<b_i^2(\alpha)\right> - \left<b_i(\alpha)\right>^2}.
\label{eqn:bias_rms}
\end{align}

The bias covariance matrices gave us the error bars on the biases at a given $\alpha$ and $c$; equation (\ref{eqn:bias_rms}) now gives us error bars on the biases at a given $\alpha$ having accounted for the full posterior in $c$.  Thus the error bar estimate from this marginalization will be greater than or equal to that from the bias covariance matrix: $\sigma(b_i(\alpha)) \geq \sigma_{b_i}(\alpha, c)$.

We do not marginalize the biases over $\alpha$. Doing so would correspond to accounting for uncertainties in the cosmological parameters when computing the biases. Generally when one fits for biases the cosmology is known to much higher precision than the biases, and indeed if one did wish to marginalize the biases over $\alpha$ here one would likely do so with an extremely restrictive prior on $\alpha$, as $\alpha$ is quite well-constrained from e.g. the BAO method in the 2PCF.  Rather, we simply select the bias and root mean square of the bias at the best-fit $\alpha$, i.e. the $\alpha$ with the minimal $\chi^2$.  

We also note that with the bias parameters in hand for each of 298 mock catalogs as well as the data, we have a third way of estimating the error bars on the biases. We can simply take the standard deviation of the best-fit biases over the 298 mocks: this is the scatter in the bias expected if we were to measure 298 realizations of the same underlying initial Gaussian-random density field.

\subsection{$\alpha$ marginalization}
\label{subsec:alpha_marginalization}
For the measurements of $\alpha$ we report, we use a similar marginalization procedure to that in \S\ref{subsec:bias_marginalization}. Here we compute
\begin{align}
\left<\alpha^n \right> = \frac{1}{N_{\alpha}(\alpha)}\int d\alpha\; dc\;\alpha^n\exp\left[-\frac{\chi^2(\alpha,c)}{2}\right]
\label{eqn:alpha_expectation_value}
\end{align}
where $N_{\alpha}$ is the normalization given by equation (\ref{eqn:alpha_expectation_value}) evaluated with $n=0$.
Again we have neglected a factor of $\det \bf{C}_{\rm biases}$ in these integrals for the same reasons as  discussed in \S\ref{subsec:bias_marginalization}.

Our marginalization procedure here results in a single value for $\left<\alpha^n\right>$ for each catalog; there is no need for further selecting the best-fit $\alpha$ as this parameter has already been integrated out. Analogously to \S\ref{subsec:bias_marginalization}, successively setting $n=1$ and $n=2$ we can obtain $\sigma(\alpha)$, the root mean square of $\alpha$.  Since, unlike the biases, $\alpha$ was not fit in our Gaussian likelihood approach, we do not have a covariance matrix error bar for $\left<\alpha\right >$.  Thus $\sigma(\alpha)$ provides an important estimate of the error in the distance scale.  We can also compute the standard deviation of $\left<\alpha\right >$ over all 298 mock catalogs for a second estimate of the precision on $\left<\alpha\right >$.

\section{BAO detection}
\label{sec:BAO_fit}
As we detailed in \S\ref{sec:fitting_procedure}, within each bias model (minimal and tidal tensor) we fit two templates to our data to assess the BAO significance. First, we fit a physical template including BAO. Second, we fit a ``no-wiggle'' template where the BAO have been removed.  The square-root of the $\chi^2$ difference between ``no-wiggle'' and physical templates gives the BAO significance.  All of our fitted parameters are summarized in Table 1 (data) and Table 2 (mocks). In Table 2 we have averaged each parameter over its value for each of the 298 mocks.

In the minimal model, we detect the BAO at $4.5\sigma$, corresponding to a $\Delta\chi^2 = 20.03$ between the best-$\alpha$ no-wiggle and BAO templates. In the tidal tensor model, we detect the BAO at $4.4\sigma$, corresponding to a $\Delta\chi^2 = 19.08$ between the best-$\alpha$ no-wiggle and BAO templates.  The average $\Delta\chi^2$ for the mocks is comparable to what we find in the data, showing that this BAO significance is typical for a survey of the given volume. 

The tidal model is a slightly better fit to the data than the minimal model, with $\chi^2/{\rm d.o.f.}=216.42/195$ as opposed to $\chi^2/{\rm d.o.f.}= 223.22/196$ for the minimal model. These $\chi^2$ values have probabilities of respectively $14.0\%$ and $8.9\%$ to occur by chance if the model is an adequate descriptor of the data. For the mocks, we find $\chi^2/{\rm d.o.f.}= 194.84/195$ for the tidal model and $\chi^2/{\rm d.o.f.}= 196.13/196$ for the minimal model. The probabilities for these $\chi^2$ to occur by chance if these models are adequate descriptors of the mocks are respectively $49\%$ and $48\%$. It is notable that the model fits the data and the mocks so well.  It is non-trivial that the $\chi^2/{\rm d.o.f.}$ for the mocks is near unity. Though the covariance was scaled to the variance of the mocks, this did not require that the actual mock 3PCF would be well-fit by the model.  It is even more notable that the data has $\chi^2/{\rm d.o.f.}$ near unity.

\begin{table}
\label{table:best_fit_params_both_data}
\begin{tabular}{|c|c|c|}
\hline
Data \tabularnewline\hline
\;\;\;\;\; & \specialcell{Minimal} &\specialcell{Tidal}\tabularnewline\hline
$\Delta\chi^2$&\specialcell{$20.03$} & \specialcell{$19.08$}\tabularnewline\hline
$\alpha$& \specialcell{$0.990\pm0.016$} & \specialcell{$0.985\pm0.014$}\tabularnewline\hline
$b_1$ &\specialcell{$1.788\pm0.018$} & \specialcell{$2.069\pm0.083$}\tabularnewline\hline
$b_2$ &\specialcell{$0.50\pm0.16$} & \specialcell{$0.08\pm0.17$}\tabularnewline\hline
$b_t$ &\specialcell{---} & \specialcell{$-0.35\pm0.14$}\tabularnewline\hline
$c$ &\specialcell{$-0.014\pm0.003$} & \specialcell{$-0.014\pm0.003$}\tabularnewline\hline
$\chi^2$&\specialcell{$223.22$} & \specialcell{$216.42$}\tabularnewline\hline
\end{tabular}
\centering
\caption{Table of best-fit parameters for the CMASS data.  $b_1$, $b_2$, and $b_t$ are the linear, non-linear, and tidal tensor biases, and $c$ encodes the integral constraint (\S\ref{sec:models}).  $\Delta\chi^2$ describes the $\chi^2$ penalty a no-BAO model (\S\ref{sec:BAO_fit}) pays over a model with BAO. The values listed here imply a $4.5\sigma$ BAO detection for the minimal model and a $4.4\sigma$ BAO detection for the tidal tensor model. $\alpha$ describes the inferred cosmic distance scale. The error bars quoted here are from the square root of the diagonal of the bias covariance matrix $\bf{C}_{\rm bias}$, as further described in \S\ref{subsec:bias_fitting}. Our error bars on the linear bias correspond to $1.0\%$ for the minimal model and $4.0\%$ for the tidal tensor model (note this is at fixed $\sigma_8$).}
\end{table}

\begin{table}
\label{table:best_fit_params_both_mocks}
\begin{tabular}{|c|c|c|}
\hline
Mocks \tabularnewline\hline
\;\;\;\;\; & \specialcell{Minimal} &\specialcell{Tidal}\tabularnewline\hline
$\Delta\chi^2$&\specialcell{$19.80$} & \specialcell{$19.91$}\tabularnewline\hline
$\alpha$& \specialcell{$1.006\pm0.02\;(0.02)$} & \specialcell{$1.008\pm0.03\;(0.02)$}\tabularnewline\hline
$b_1$ &\specialcell{$1.900\pm0.029$} & \specialcell{$1.573\pm0.115$}\tabularnewline\hline
$b_2$ &\specialcell{$0.48\pm0.20$} & \specialcell{$0.66\pm0.33$}\tabularnewline\hline
$b_t$ &\specialcell{---} & \specialcell{$0.13\pm0.25$}\tabularnewline\hline
$c$ &\specialcell{$0.000\pm0.009$} & \specialcell{$0.000\pm0.009$}\tabularnewline\hline
$\chi^2$&\specialcell{$196.13$} & \specialcell{$194.84$}\tabularnewline\hline
\end{tabular}
\centering
\caption{Table of best-fit parameters for the mocks. We report the mean of each parameter over the 298 mocks. The error bars on $\alpha$ are the standard deviation of the marginalized $\left <\alpha\right>$ over the 298 mocks, and in parentheses we also report the average of the root mean square $\sigma(\alpha)$ over the 298 mocks. For the error bars on the bias parameters and $c$ we hold $\alpha$ fixed at its mean over all the mocks, as further discussed in the main text. Comparing the error bars here, from the scatter of the 298 mocks, to those reported in Table 1, mostly confirms that the error bars estimated from the bias covariance matrix are reasonable.  We further discuss this point in \S\ref{sec:bias_parameters}.}
\end{table}

\begin{figure*}
\centering
\includegraphics[width=.48\textwidth]{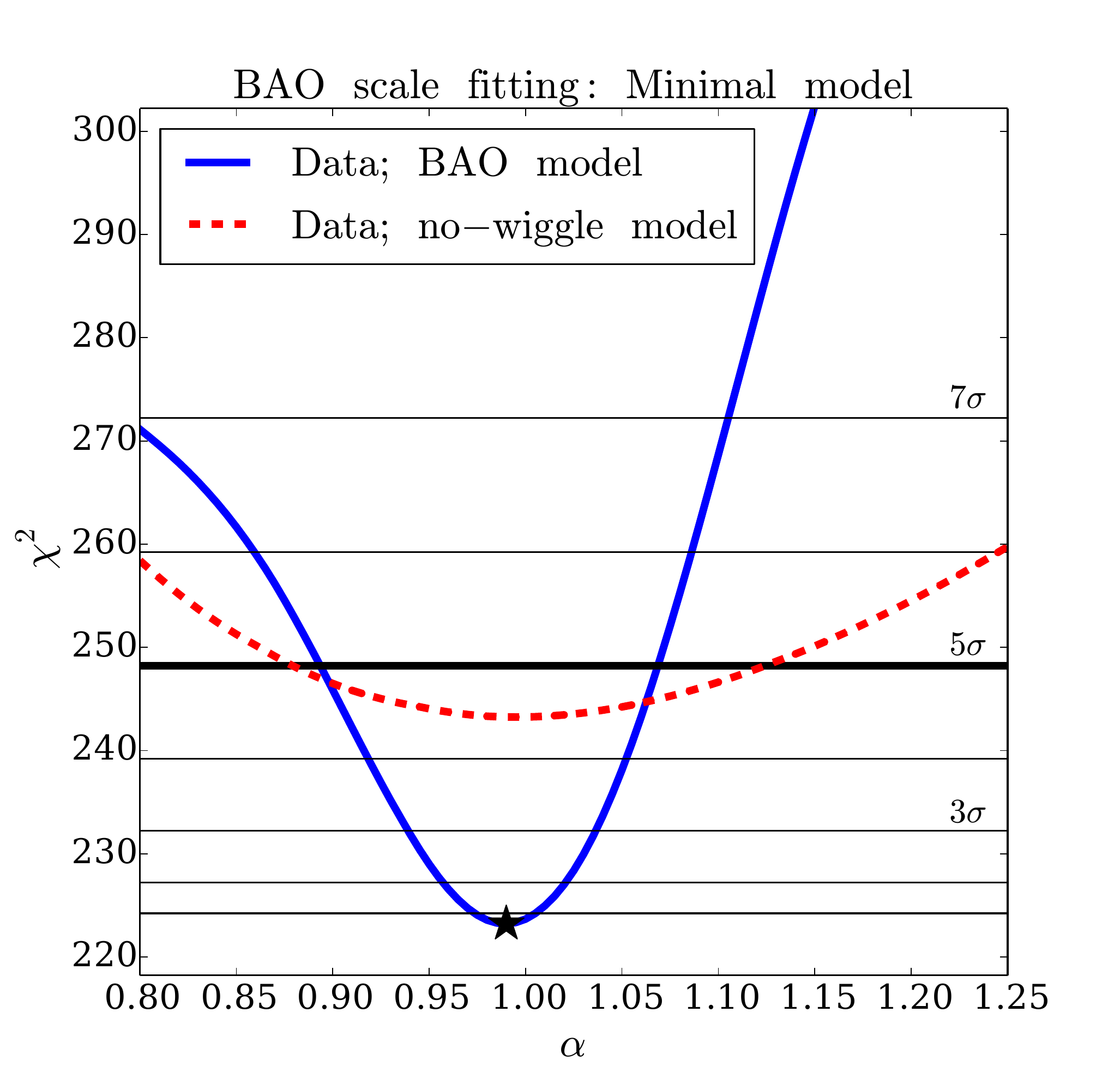}
\includegraphics[width=.51\textwidth]{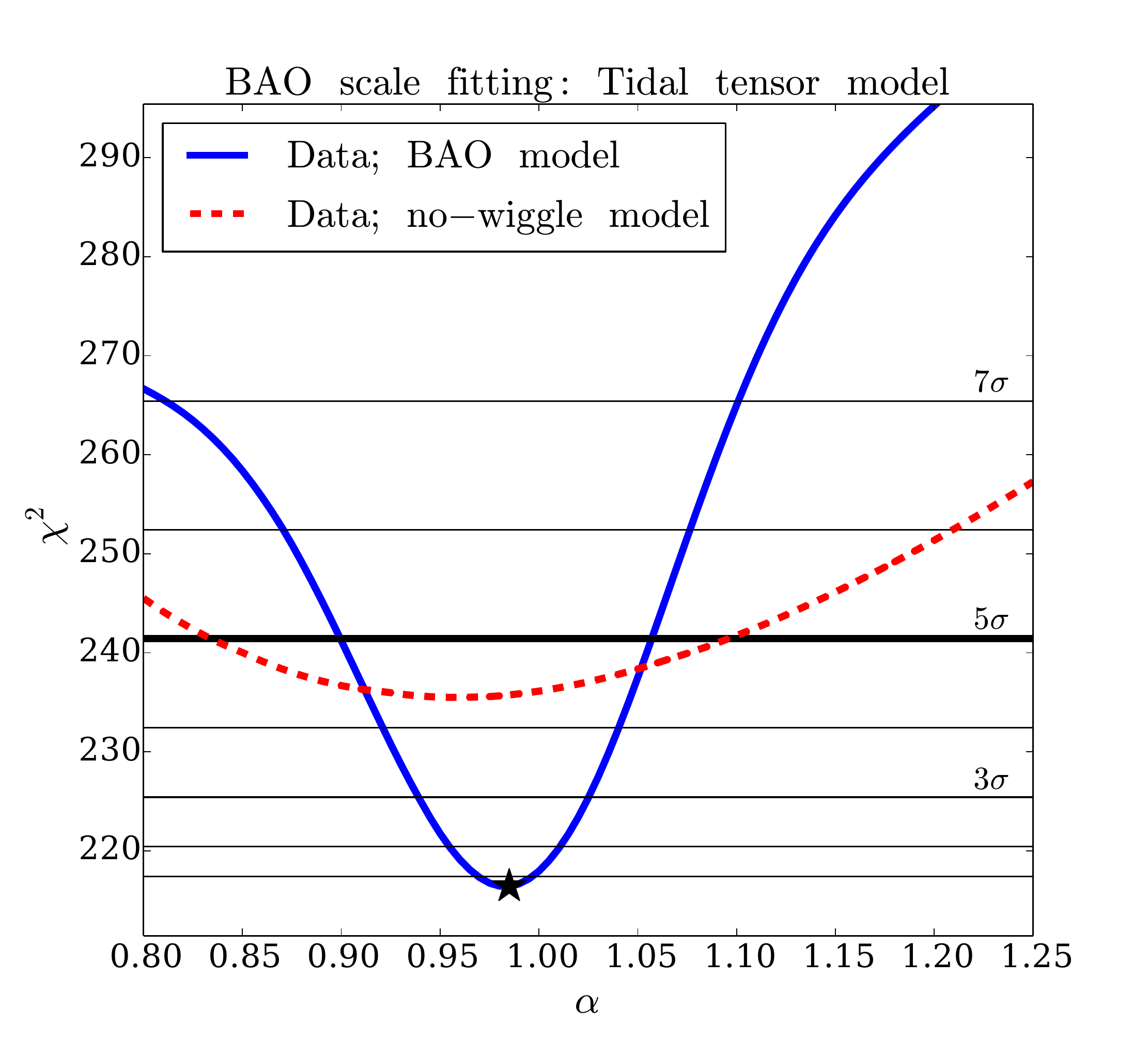}
\includegraphics[width=.52\textwidth]{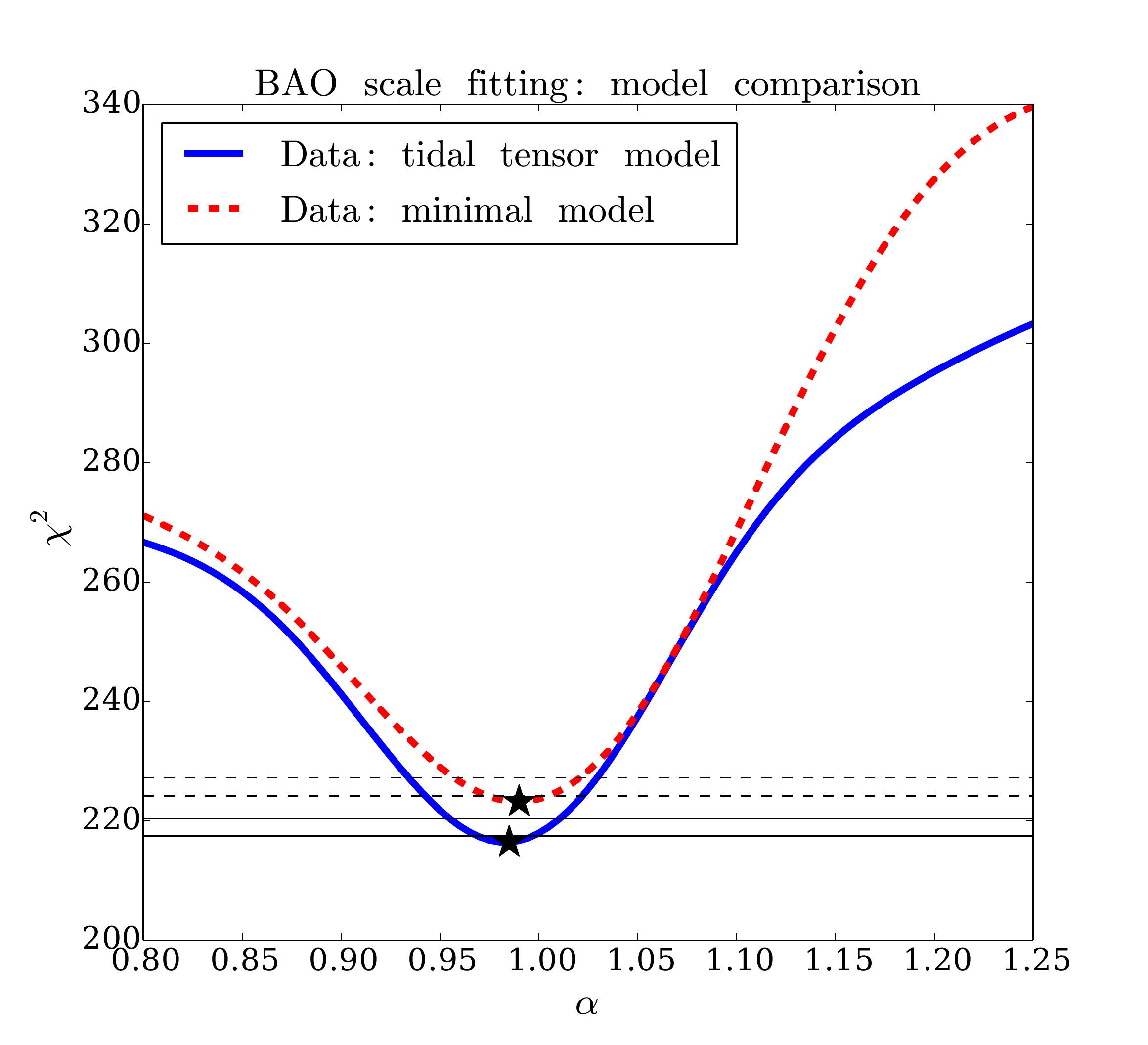}
\caption{The upper panels show the best-fit BAO and no-wiggle models for the data vs. the distance scale parameter $\alpha$. For each, we have indicated the best-fit $\alpha$ with a black star. In both models the best-fit BAO template is preferred at roughly $4.5\sigma$ to the best-fit no-wiggle template. The lower panel shows the BAO templates for each bias model, with best-fit $\alpha$ again denoted by stars. The horizontal lines in this lower panel denote $1\sigma$ and $2\sigma$ thresholds for each model, solid for tidal tensor and dashed for minimal. The tidal tensor model provides a slightly better fit to the data, and both $\chi^2$ curves have similar widths with respect to $\alpha$, suggesting our distance scale precision should be robust to bias model choice. Further discussion of these plots is in \S\ref{sec:BAO_fit}.}
\label{fig:BAO_scale_chisq}
\end{figure*}

Figure \ref{fig:BAO_scale_chisq} further explores our BAO detection.  Within each model, the best-fit $\alpha$ for the no-wiggle template and physical template is nearly the same. $\alpha$ is free within each template, and so this similarity of the best-fit $\alpha$ is by chance and did not have to be the case. These plots shows that relative to the best-fit no-wiggle model, the best-fit BAO models live in roughly $4.5\sigma$  valleys.  The similarity of the $\chi^2$ valley in both upper panels indicates that the BAO detection is robust to bias model choice. While the physical template is $4.5\sigma$ better than the no-wiggle template, in fact within the physical template the rejection of alternative $\alpha$'s has a much steeper divot than this: we reject alternate values of $\alpha$ at roughly $7\sigma$.  The no-wiggle template is an interesting null hypothesis only for testing for the BAO's presence. Once the BAO are assumed, the steep divot rejecting alternate values of $\alpha$ permits a highly precise constraint on the cosmic distance scale.

The best-fit $\alpha$ for the physical templates within each model is indicated with a black star.  The narrowness of the $\chi^2$ valley with respect to $\alpha$ indicates that we should find a very precise constraint on the cosmic distance scale from these BAO detections; we will return to this point in \S\ref{sec:distance_scale}.  In the lower panel of Figure \ref{fig:BAO_scale_chisq}, we show both minimal and tidal tensor models for the physical power spectrum template only to permit comparison of these two models.  Again we indicate the best-fit $\alpha$ for each model with a black star.  This lower panel also shows that the tidal tensor model is overall a slightly better fit to the data than the minimal model, as its minimal $\chi^2$ is lower.  The similar width about their respective minima of the $\chi^2$ curves in the lower panel shows that the precision of the constraint on $\alpha$ is also robust to bias model choice.

Overall, there is mild evidence that a tidal tensor bias is required. From Table 1, $\Delta\chi^2=6.80$ between the tidal tensor model with physical template and the minimal model with physical template, meaning a $2.6\sigma$ preference for tidal tensor bias.

\begin{figure*}
\centering
\includegraphics[width=.48\textwidth]{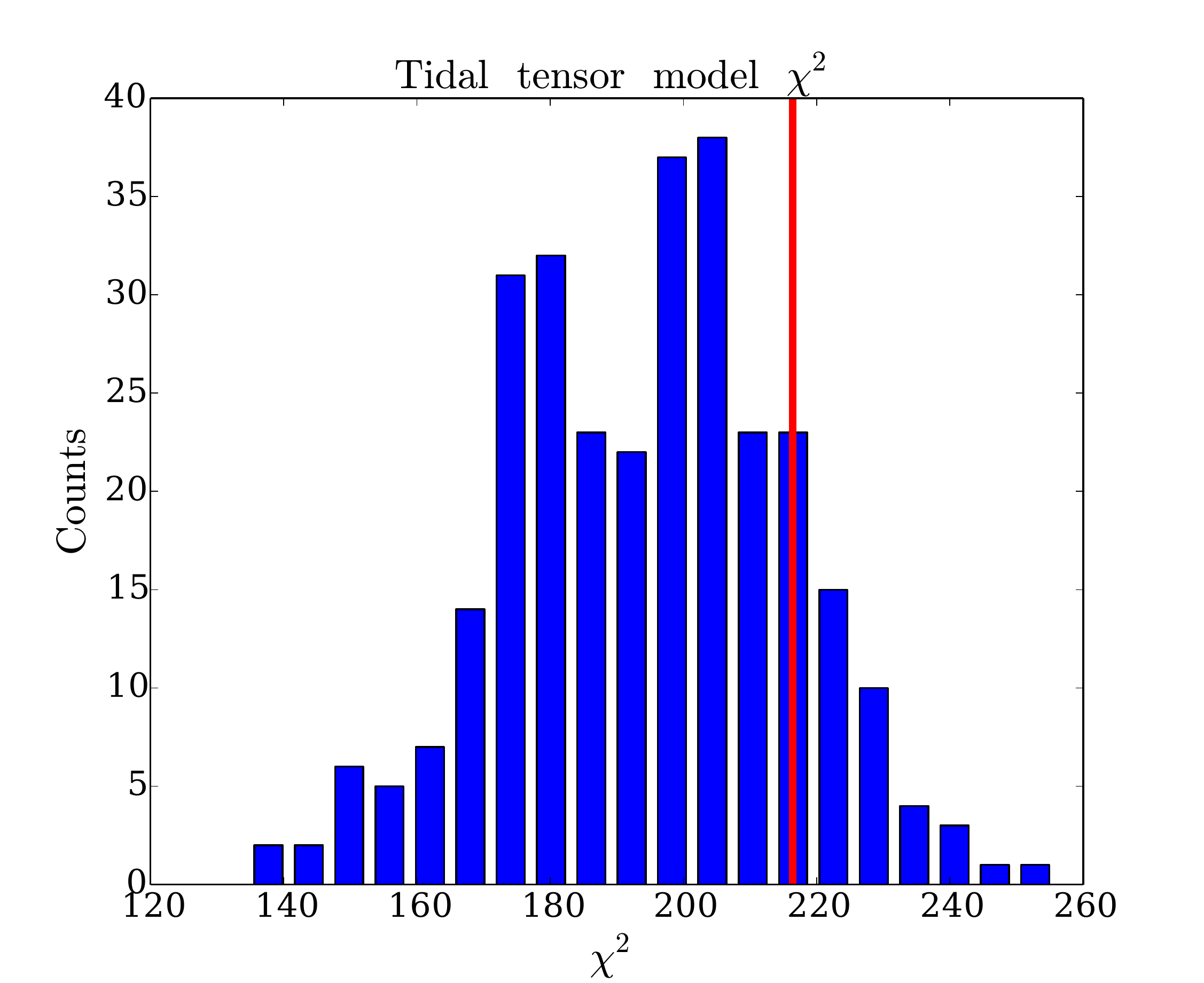} 
\includegraphics[width=.515\textwidth]{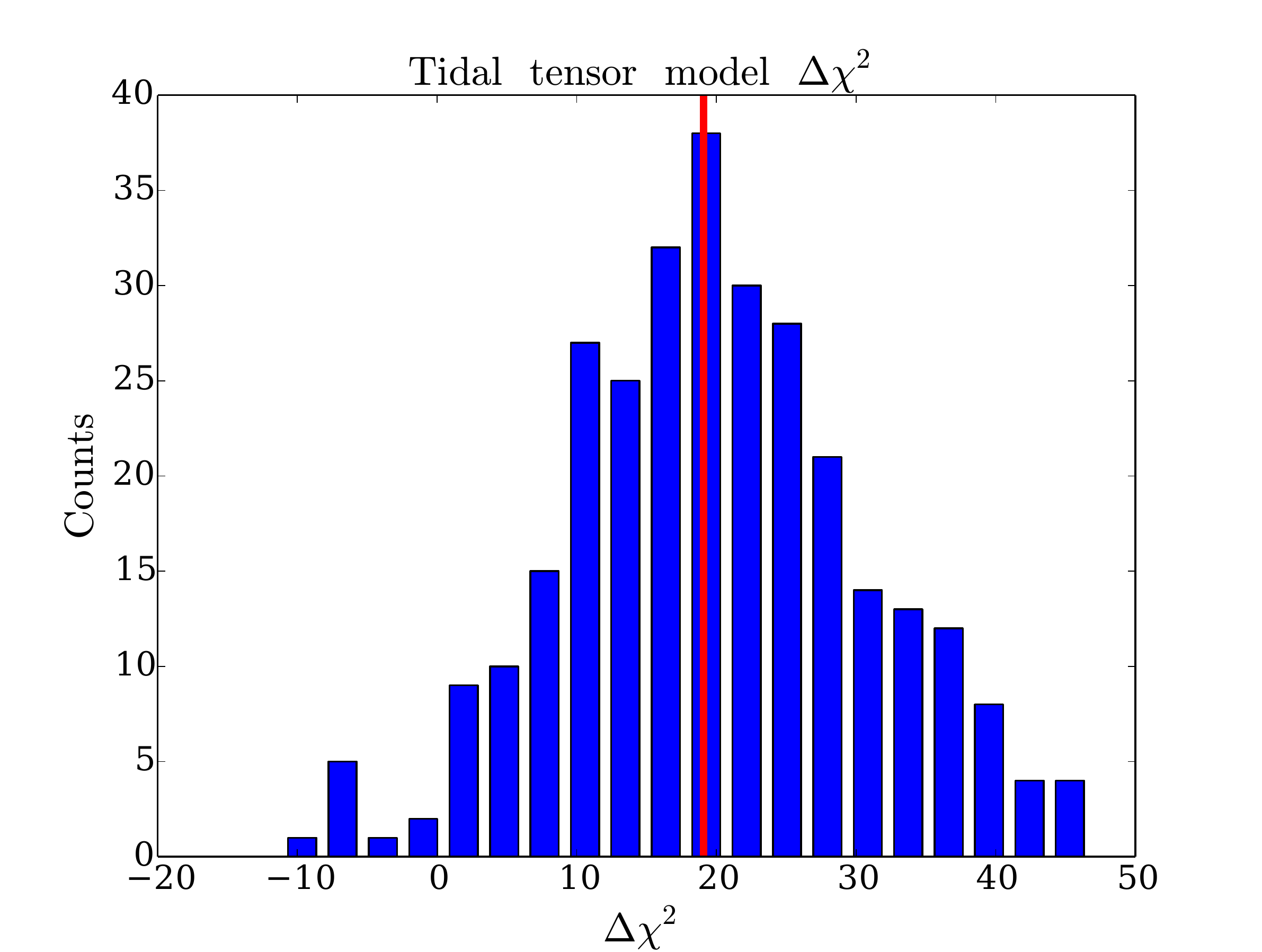} 
\includegraphics[width=.48 \textwidth]{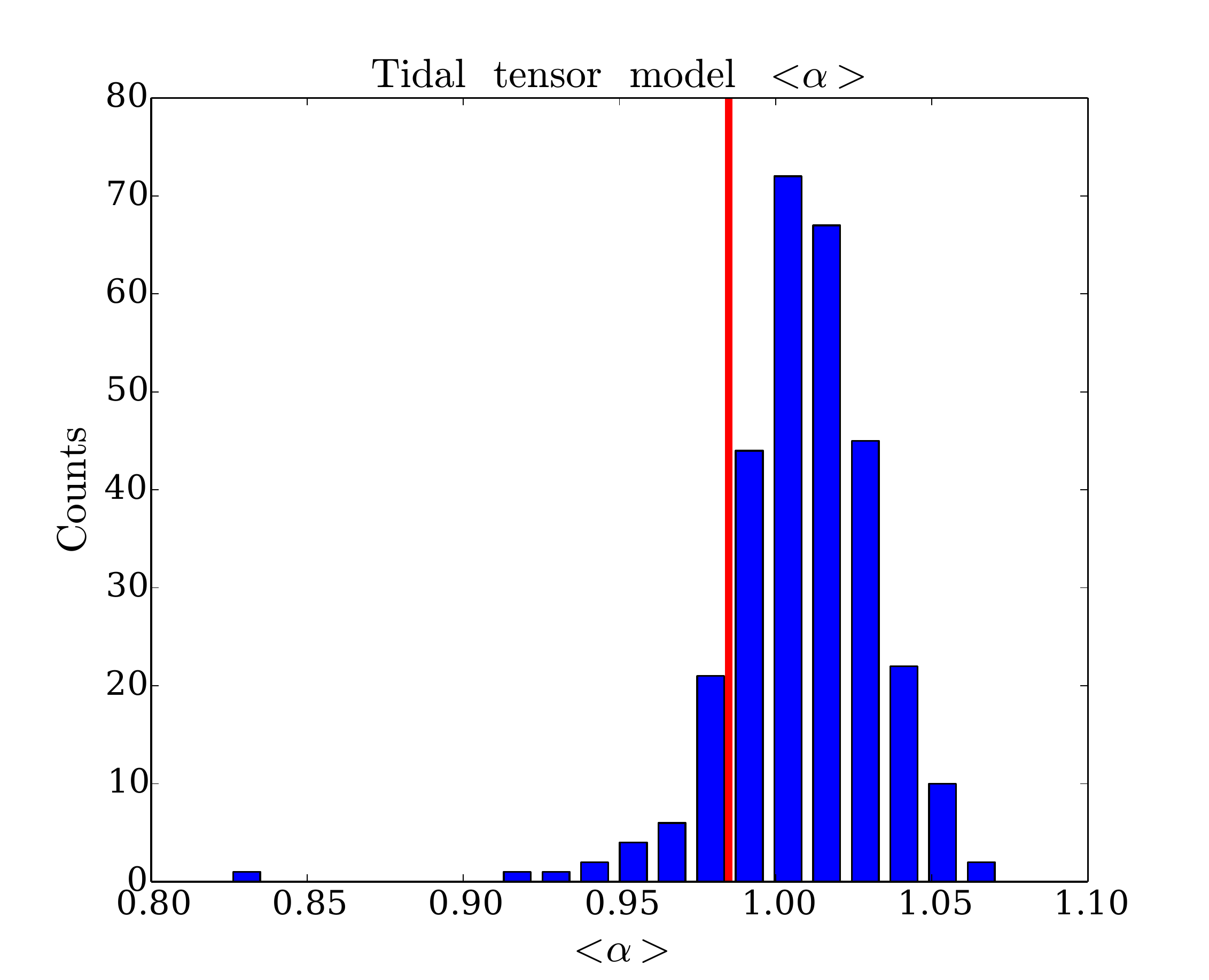} 
\includegraphics[width=.51\textwidth]{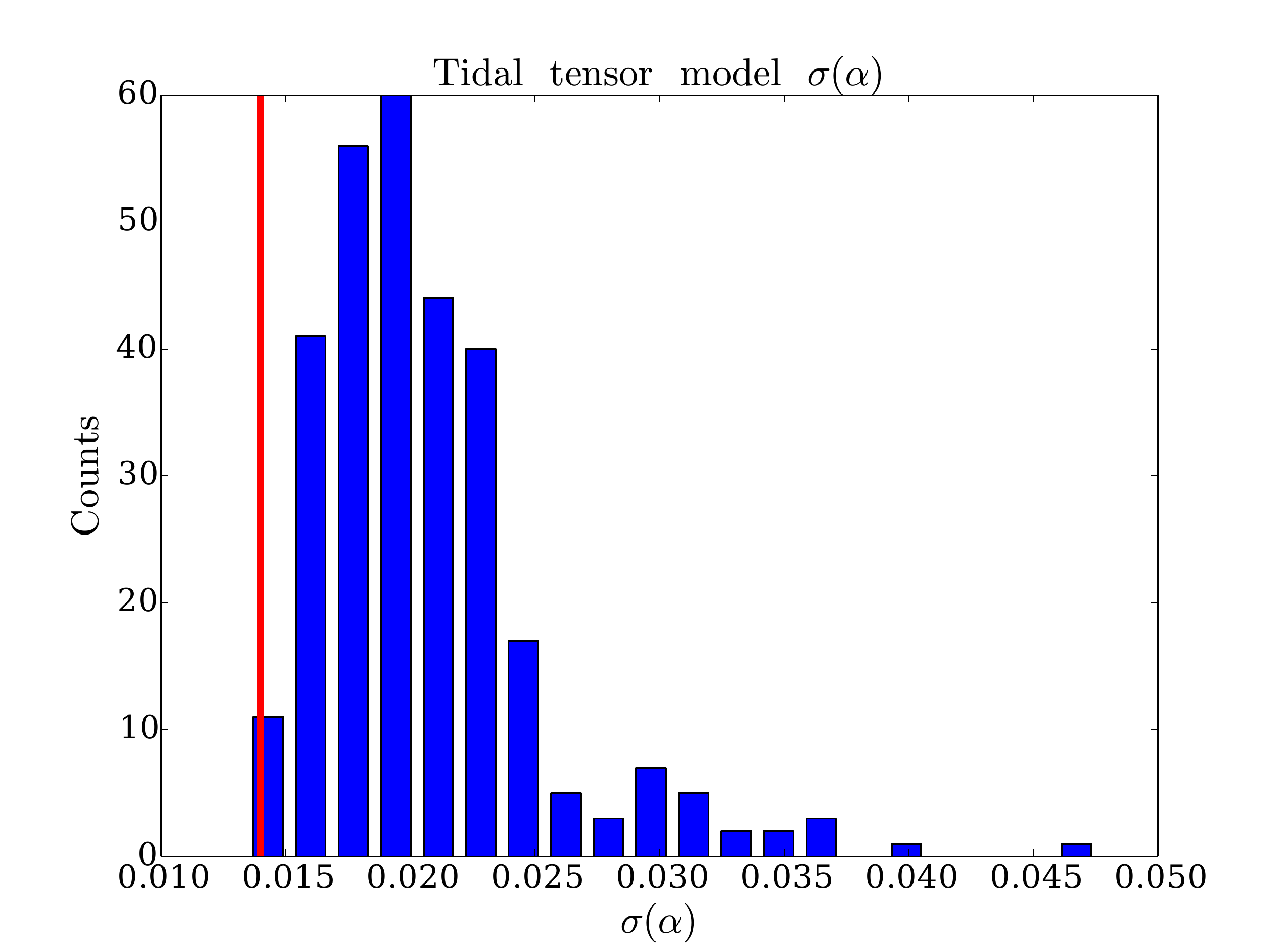} 
\caption{The upper two panels show histograms of the best-fit $\chi^2$ and the $\Delta\chi^2$ with respect to the best-fit no-wiggle templates for the 298 {\textsc PATCHY} mocks, with the red vertical line indicating the data values. These show that our goodness of fit and BAO significance are both fairly typical for a survey of this volume.  The bottom panels show histograms of the mock results for the marginalized $\left<\alpha \right>$ and root mean square $\sigma(\alpha)$.  The fiducial cosmology used for fitting agrees with that for the mocks, so $\left<\alpha \right>$ should center at unity. However, there is a $0.8\%$ offset between unity and the mean $\left<\alpha \right>$ over all mocks; this possible systematic is discussed further in \S\ref{sec:BAO_fit}. The scatter in $\left<\alpha \right>$ is about $2\%$, consistent with the typical root mean square shown in the right panel. This right panel shows that our precision on the distance scale is better than most of the mocks but comparable to that from the best ten mocks.}

\label{fig:tidal_first_four}
\end{figure*}

The top two panels of Figure \ref{fig:tidal_first_four} illustrate that our results are typical given the survey volume and the tidal tensor bias model. The left panel shows a histogram of the $\chi^2$ for 298 mocks and with the data value marked as a red vertical line, shows that our best-fit $\chi^2$ is fairly typical. The right panel shows a histogram of the $\Delta\chi^2$ relative to the best-fit no-wiggle template, shows that our BAO detection significance is also fairly typical.  

The bottom two panels of Figure \ref{fig:tidal_first_four} show the distance scale mean $\left<\alpha \right>$ and root mean square $\sigma(\alpha)$ for the mocks and data; again the data values are indicated by red lines.  As both Table 2 and the bottom left panel reveal, the mean $\alpha$ for the mocks is shifted by $0.9\%$ relative to unity. If our estimator for $\alpha$ is unbiased, we should recover $\left<\alpha \right> = 1$ averaged over all of the mocks, since we knew the correct cosmology for the mocks.  It is not clear whether this bias is intrinsic to the 3PCF as an estimator of $\alpha$, whether it reflects some undiagnosed issue with the mocks themselves, or whether it is an actual shift due to non-linear structure formation or bias. 

\begin{figure*}
\centering
\includegraphics[width=.48\textwidth]{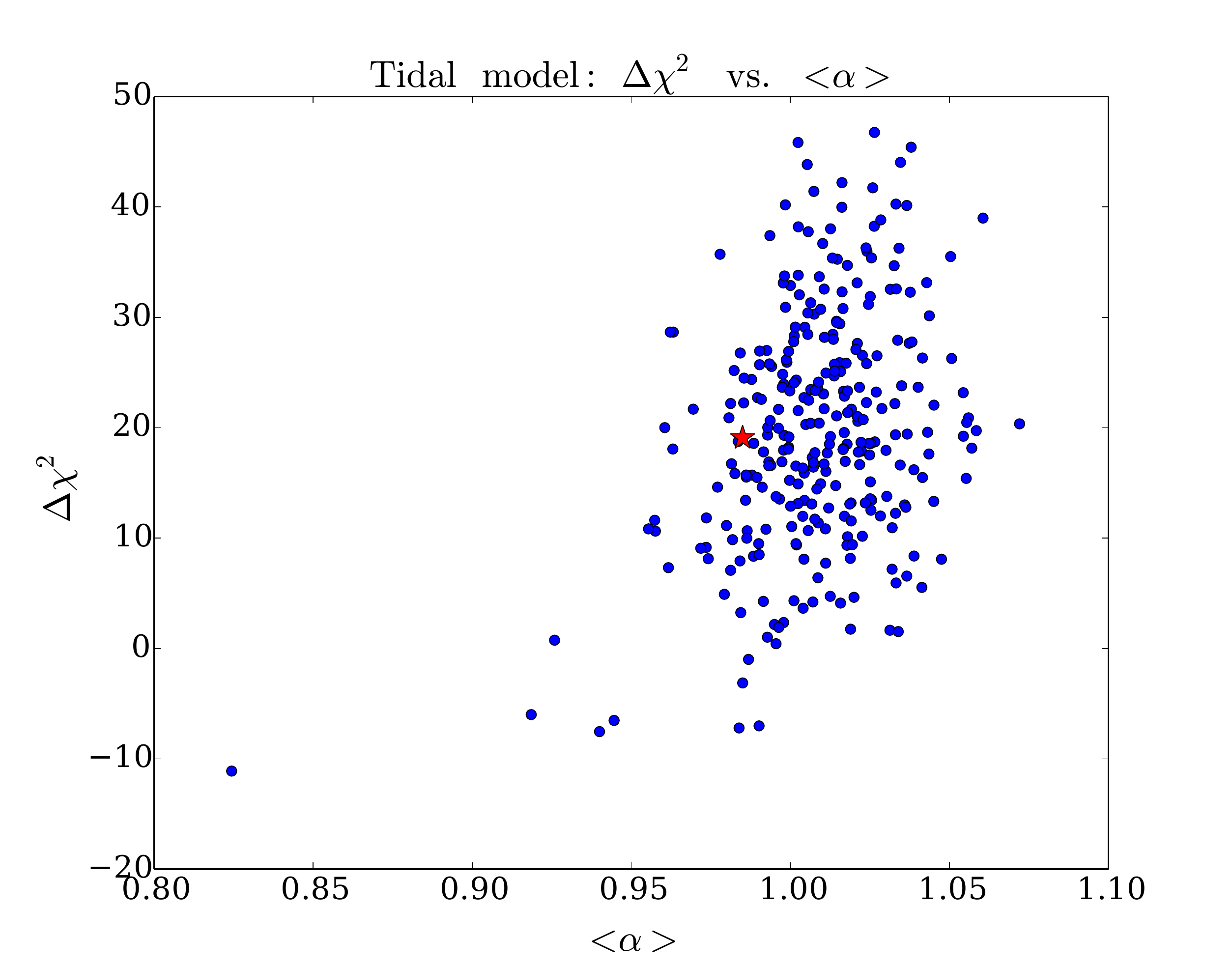}
\includegraphics[width=.51\textwidth]{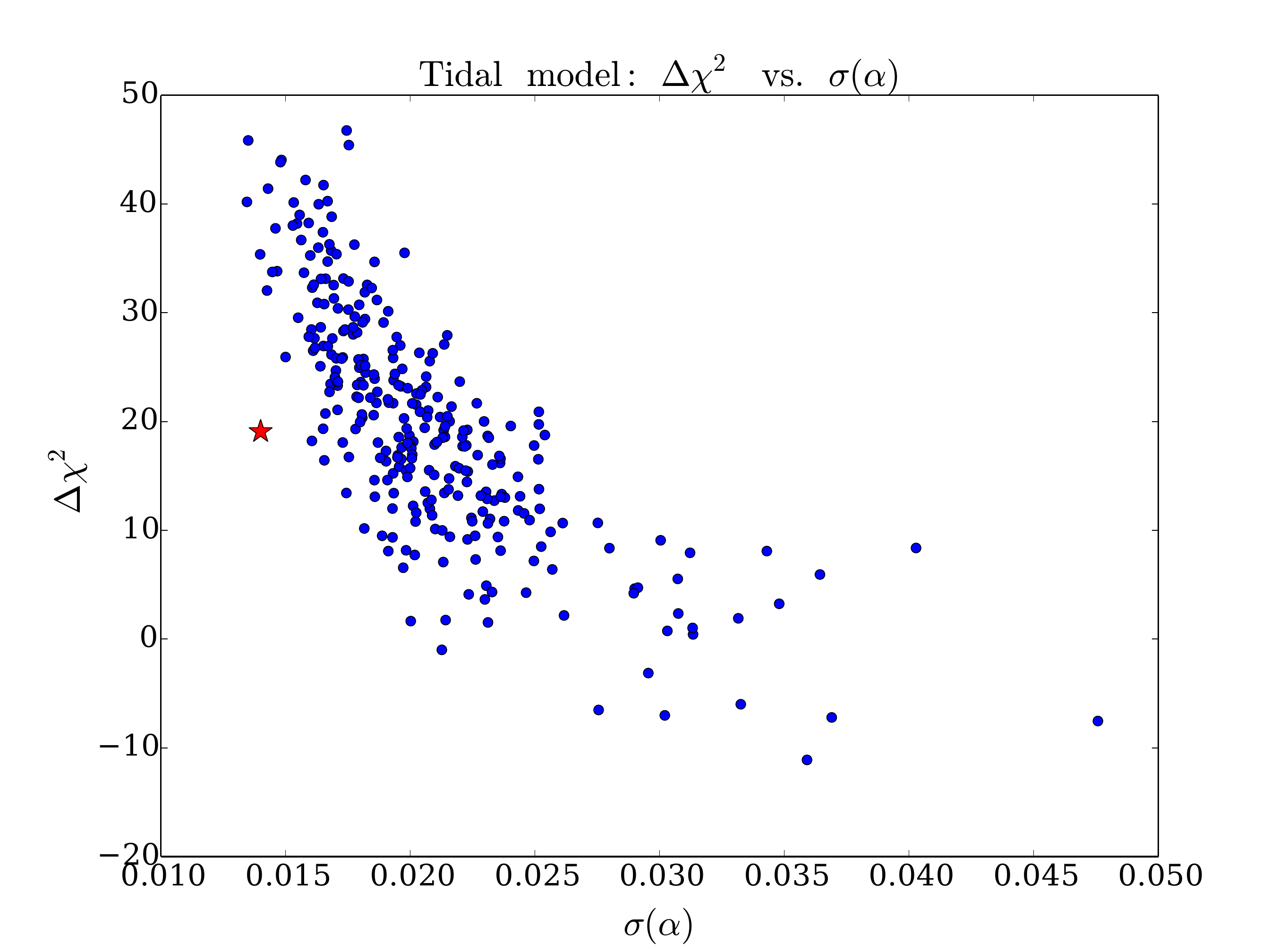}
\caption{Blue points denote mocks and the red star denotes data. In the left panel we show the $\Delta\chi^2$ relative to the best fit no-wiggle template for the tidal model. The significance of our BAO detection is not highly correlated with $\left<\alpha \right>$. In the right panel we show the detection significance versus the root mean square $\sigma(\alpha)$.  As expected, the stronger our detection the smaller the error on the distance scale.}
\label{fig:scatter_delta_chisq}
\end{figure*}

Figure \ref{fig:scatter_delta_chisq}, left panel, shows the difference in $\chi^2$ between the best-fit no-wiggle template and the best-fit physical template for the tidal tensor model for 298 mocks (blue points) and the data (red star).  This $\Delta\chi^2$ is plotted versus the marginalized $\alpha$.  This panel illustrates that the significance of our BAO detection is not highly correlated with $\left< \alpha \right>$. Figure \ref{fig:scatter_delta_chisq}, right panel, shows $\Delta\chi^2$ versus $\sigma(\alpha)$, the root mean square error on $\alpha$.  As expected the stronger the BAO detection the smaller $\sigma(\alpha)$. Our data value (red star) has a somewhat better precision on $\alpha$ than expected from the mocks given the $\Delta\chi^2$.  

It is well-known that for the 2PCF order half-percent shifts in $\left<\alpha \right>$ as estimated from mock catalogs are possible. In future work we will explore what the correct treatment is for this possible systematic in $\alpha$ as measured from the 3PCF. At present we conservatively elect to incorporate an additional $1\%$ systematic error in our reported precision on $\alpha$ as measured from the data.

\section{Cosmic Distance scale}
\label{sec:distance_scale}
\subsection{Measured $\DV$ and comparison with other works}
To convert $\alpha$ into a physical distance scale $\DV$ to redshift $0.57$, we generalize the formula for $\DV$ of Anderson et al. (2014) to varying $\omegam$ and redshift; we also convert to the \textsc{Patchy} cosmology and from $\Mpch$ to ${\rm Mpc}$. We find
\begin{align}
\DV(z_{\rm survey}) = \alpha \times 2054.4\;{\rm Mpc}\left(\frac{r_{\rm d}}{r_{\rm d,\;{\textsc PATCHY}}}\right)
\end{align}
where $r_{\rm d}$ is the sound horizon at decoupling, $r_{\rm d,\;{\textsc PATCHY}} = 147.66\;{\rm Mpc}$ is the sound horizon at decoupling for the \textsc{Patchy} cosmology, and $z_{\rm survey} = 0.57$. We thus find
\begin{align}
&D_{\rm V,\; minimal}(z_{\rm survey})  = 2034\pm 33\;{\rm Mpc\;(stat)}\left(\frac{r_{\rm d}}{r_{\rm d,\;{\textsc PATCHY}}}\right),\nonumber\\
&D_{\rm V,\;tidal}(z_{\rm survey})  = 2024\pm 29\;{\rm Mpc\;(stat)}\left(\frac{r_{\rm d}}{r_{\rm d,\;{\textsc PATCHY}}}\right).
\end{align}
For the data catalog only, we used $\omegam = 0.31$, so the $\alpha$'s have been adjusted by $1.0015$ when converting to the $\DV$ above quoted in terms of the \textsc{PATCHY} sound horizon.

From fitting the 2PCF of SDSS DR11 including reconstruction, Anderson et al. (2014) found 
\begin{align}
&D_{\rm V,\; Anderson} (z_{\rm survey}) = 2034\pm 20\;{\rm Mpc}\left(\frac{r_{\rm d}}{r_{\rm d,\;{\textsc PATCHY}}}\right)
\end{align}
while from the SDSS DR12 CMASS 2PCF including reconstruction, Cuesta et al. (2016) found
\begin{align}
&D_{\rm V,\; Cuesta}(z_{\rm survey})  = 2036\pm 21\;{\rm Mpc}\left(\frac{r_{\rm d}}{r_{\rm d,\;{\textsc PATCHY}}}\right)
\end{align}
From the reconstructed multipoles of the CMASS DR12 power spectrum, Gil-Mar\'in et al. (2016) found 
\begin{align}
&D_{\rm V,\; Gil-Marin}(z_{\rm survey})  = 2023\pm 18\;{\rm Mpc}\left(\frac{r_{\rm d}}{r_{\rm d,\;{\textsc PATCHY}}}\right)
\end{align}
We have adjusted the measured $\DV$'s of these works appropriately to be quoted in terms of our fiducial \textsc{Patchy} sound horizon.

Both of our distance scales are within at most $0.5\sigma$ of the Anderson et al. (2014), Cuesta et al. (2016), and Gil-Mar\'in et al. (2016) measurements.  Our slightly larger error bars reflect that we achieve a precision of roughly $1.4\%$ (statistical) on the distance scale while the 2PCF measurements achieve a precision of roughly $1.0\%$. Our results are also consistent with the measured distance scale in the final cosmological analysis of the SDSS DR12 combined sample (Alam et al. 2016).

Given the mild $\chi^2$ preference for the tidal tensor model, we choose to report our final measurement of $\DV$ that for the tidal tensor model.  Given the $1\%$ offset of the mocks' mean $\alpha$ from unity, we incorporate a $1\%$ systematic error in our error budget, finding
\begin{align}
D_{\rm V,\; S16}  = \;&2024\pm 29\;{\rm Mpc\;(stat)}+20\;{\rm Mpc\;(sys)}\nonumber\\
&\times\left(\frac{r_{\rm d}}{r_{\rm d,\;{\textsc PATCHY}}}\right)
\end{align}

\begin{figure}
\centering
\includegraphics[width=.52\textwidth]{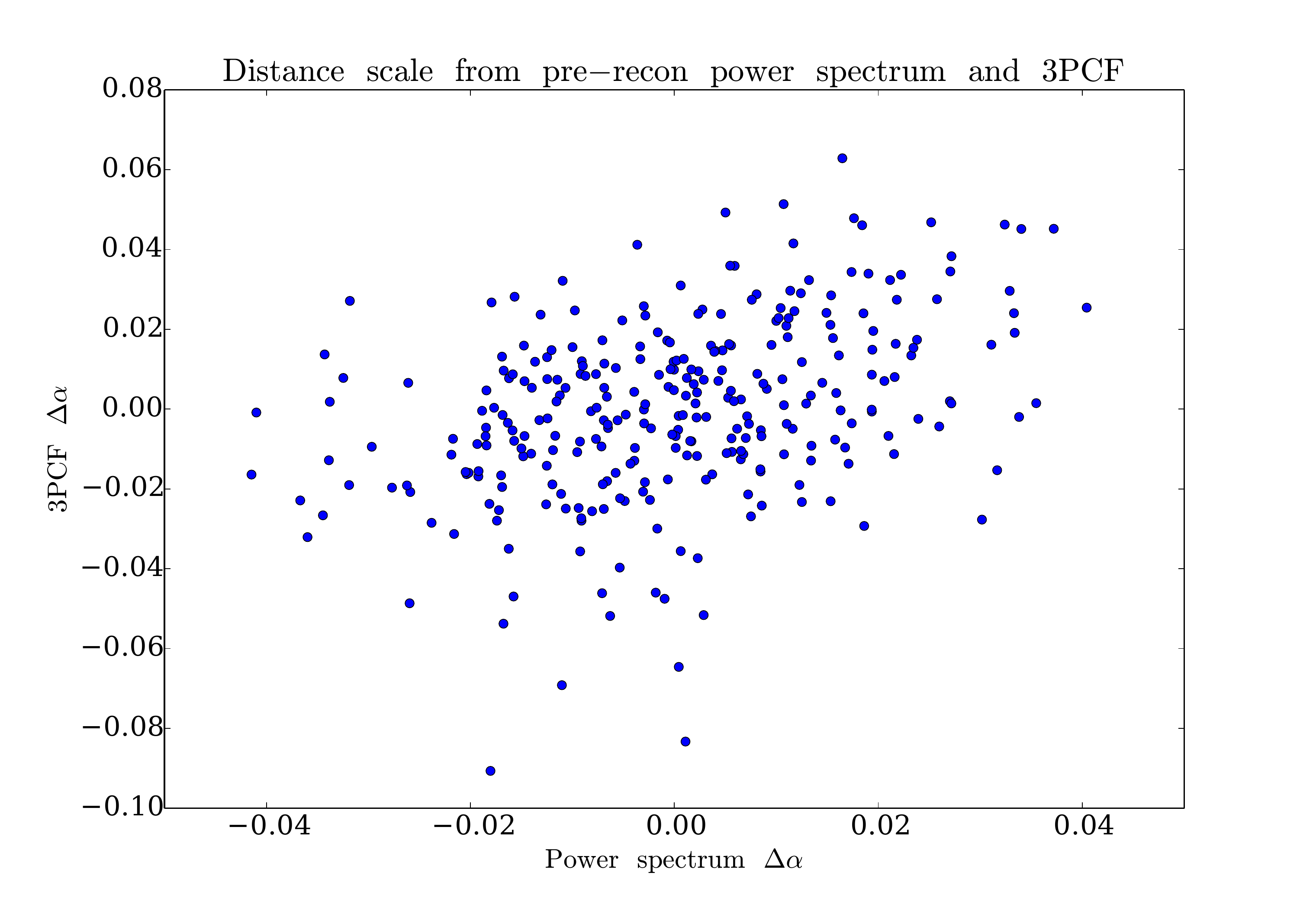}
\caption{Here we show for 297 mocks $\Delta\alpha\equiv \alpha -1$ for the 3PCF versus this quantity for the power spectrum; both are prior to density field reconstruction. We have removed mock \#292 as an outlier from this plot; its 3PCF $\Delta\alpha$ is of order $-0.2$. The correlation coefficient of the plotted points is $0.39$.  The 3PCF $\Delta\alpha$ has larger intrinsic scatter than the power spectrum $\Delta\alpha$, which can reduce the measured correlation. As discussed in the main text, the maximum possible correlation between the $\Delta\alpha$ given the additional scatter in the 3PCF is $0.71$. That the measured value of $0.39$ is well smaller than this indicates that there is new BAO information in the 3PCF.}
\label{fig:alpha_2pcf_3pcf}
\end{figure}

\subsection{Independence from the 2PCF and power spectrum}
We now address the independence of the distance information in the 2PCF and 3PCF. It has recently been shown (Schmittfull et al. 2015) that reconstruction (Eisenstein et al. 2007) introduces some 3PCF and 4PCF information into the 2PCF, presumably reducing the independence of the distance scales measured from the reconstructed 2PCF and from the 3PCF.\footnote{Reconstruction uses galaxy positions today as a proxy for the gravitational potential of the matter density and moves galaxies backwards along straight lines (their Zel'dovich approximation trajectories, inferred from this potential) to undo some non-linear structure formation and sharpen the BAO features.}

Prior to reconstruction, the distance scales measured from the power spectrum (roughly equivalent to the 2PCF for our purposes) and from the 3PCF appear to be fairly independent. These power spectrum measurements were made on the same \textsc{Patchy} mock catalogs used in our 3PCF analysis. In Figure \ref{fig:alpha_2pcf_3pcf}, we show the difference of each mock catalog's $\alpha$ from the mean over all catalogs for both power spectrum and 3PCF, i.e. $\Delta\alpha \equiv \alpha -\left< \alpha \right>$. Removing one outlier (not shown in Figure \ref{fig:alpha_2pcf_3pcf}), we find a correlation coefficient of $0.39$. 

However, the 3PCF's $\Delta\alpha$ have more intrinsic scatter than the power spectrum's, which can reduce the measured correlation. As a comparison, we ask what the correlation would be if the measured 3PCF $\alpha$ were simply the power spectrum $\alpha$ plus some uncorrelated Gaussian noise---in other words, if the 3PCF had no additional distance information. 
 
To generate test, ``fake''  3PCF $\alpha$ for this case, we add noise $\sigma_{\rm add}$ to the power spectrum $\alpha$ as $\sigma^2_{\rm add} = \sigma^2_{3PCF,\Delta\alpha} - \sigma^2_{P(k),\Delta\alpha}$, where $\sigma^2_{3PCF,\Delta\alpha}$ is the variance of the true 3PCF $\alpha$ and $\sigma^2_{P(k),\Delta\alpha}$ that for the power spectrum $\alpha$. We then compute the correlation between our ``fake'' 3PCF $\Delta\alpha$ and the power spectrum $\Delta\alpha$, finding $0.71\pm 0.02$.\footnote{The value and error bar represent the mean and rms of 100 realizations of the noise for each mock}  This value is the maximal possible correlation between 3PCF and power spectrum if they contained the same information on the distance scale.  The fact that our measured correlation of $0.39$ is about half this maximal possible value suggests that prior to reconstruction the 3PCF does contain significant additional information on the distance scale.

We also examine correlations between the 3PCF $\alpha$ and the 2PCF $\alpha$ after reconstruction. Only the 2PCF measurements here use the reconstructed density field; the 3PCF measurements are still for the unreconstructed density field as throughout this work. Here we find a correlation coefficient of $0.40$, to be compared to a maximum possible correlation coefficient of $0.45\pm 0.04$, computed in the same way using ``fake'' 3PCF data generated from the 2PCF $\alpha$ plus the appropriate-amplitude uncorrelated Gaussian noise.  The similarity of these values suggests that reconstruction moves nearly all the distance information in the 3PCF into the 2PCF.  

It can be shown that the optimal, variance-minimizing combination of $\alpha$ from 2PCF (or power spectrum) and 3PCF is given by using weights $\vec{w} = \boldsymbol{C}^{-1}\vec{E}/[\vec{E}^{\rm T} \boldsymbol{C}^{-1} \vec{E}]$, where $\boldsymbol{C}$ is the covariance matrix between the $\Delta\alpha$ from the 2PCF/power spectrum and the 3PCF and $\vec{E} = (1,1)$.  The combined $\alpha$ computed using these weights will have variance $1/[\vec{E}^{\rm T}\boldsymbol{C}^{-1}\vec{E}]$. 

For the pre-reconstruction $\alpha$, this combination offers an $12\%$ improvement in the variance over measuring the distance from the power spectrum alone, corresponding to lengthening BOSS from $4.5$ to $5$ years. Post-reconstruction, the improvement in variance relative to using the 2PCF alone is $0.32\%$, again suggesting that reconstruction adds nearly all the distance information in the 3PCF into the 2PCF.

\section{Bias parameters}
\label{sec:bias_parameters}
Here we briefly discuss the bias parameter values found for the mocks and for the data.  As Tables 1 and 2 show, the bias values and error bars we find are generally consistent between mocks and data, save for the tidal bias, which for the data is $b_t=-0.35$ while for the mocks it is on average $b_t = 0.13$.  The $b_t$ for the data matches well the prediction of local Lagrangian biasing $b_t = -(2/7)[b_1-1]=-0.31$ for the best-fit data value of $b_1=2.069$.  However, the mocks' value of $b_t$ has the wrong sign to fit this prediction. The \textsc{PATCHY} mocks do not include nonlocal bias terms, so it is not unexpected that we do not recover the theoretically predicted-value of $b_t$.

\begin{figure*}
\centering
\includegraphics[width=.52\textwidth]{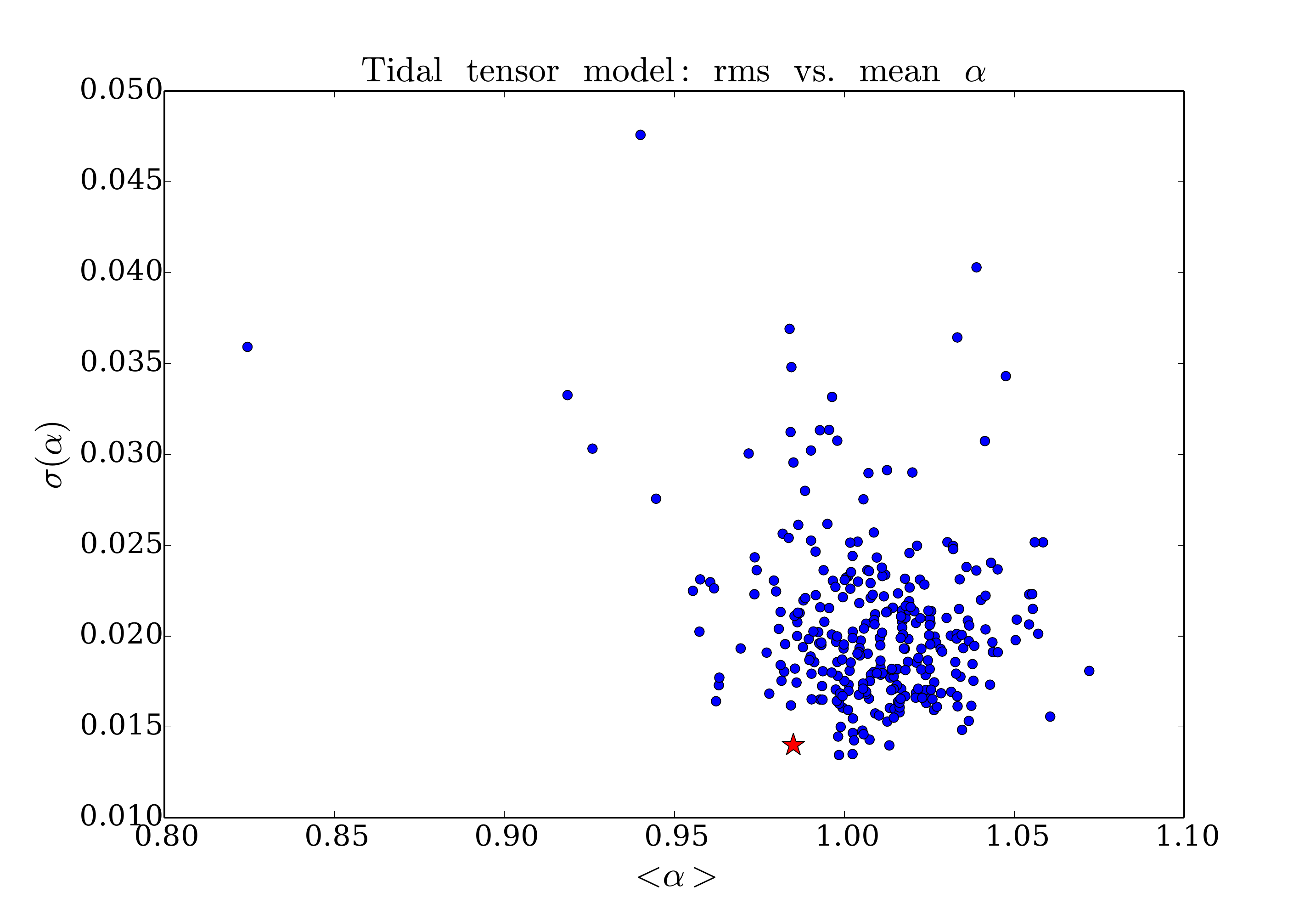}
\includegraphics[width=.47\textwidth]{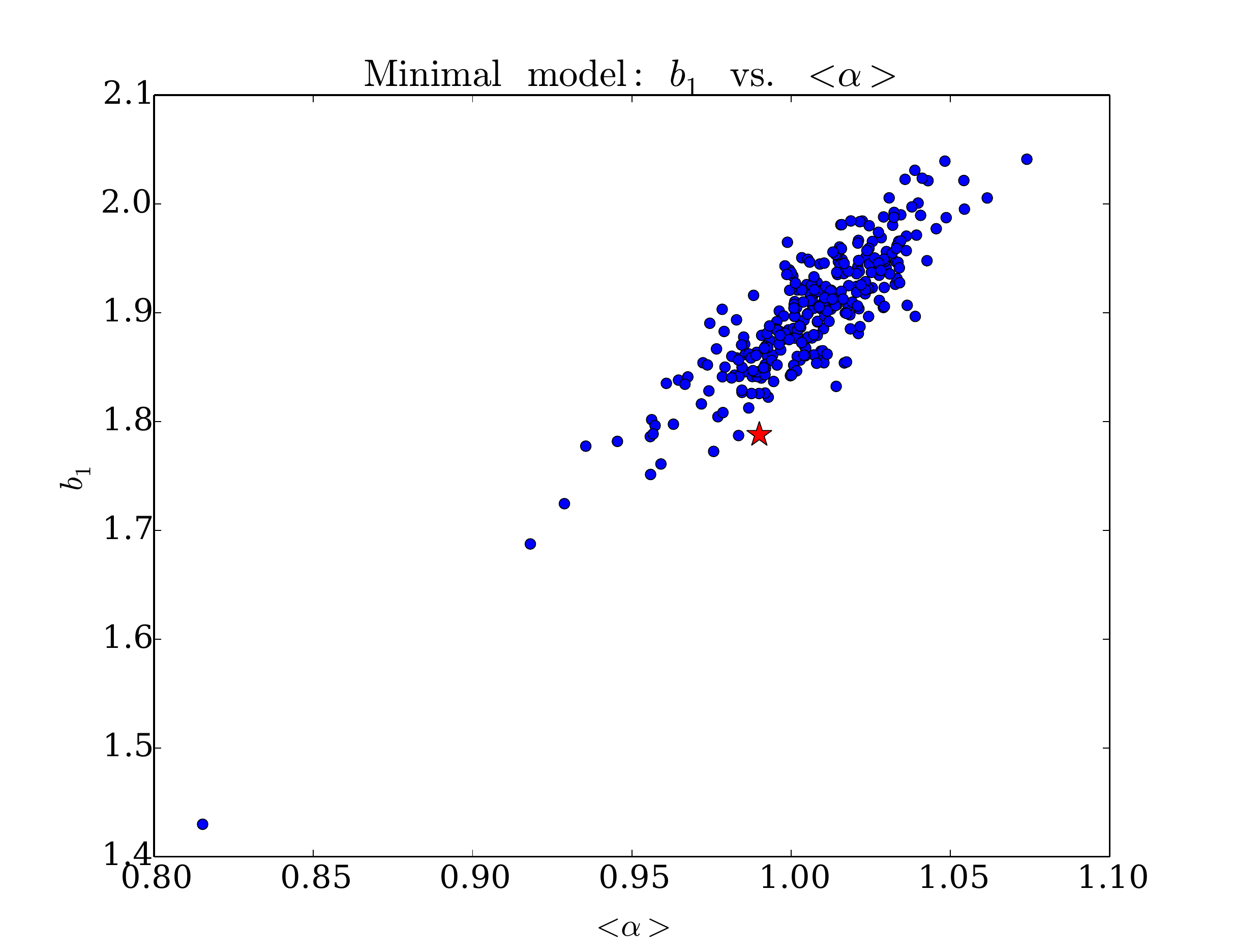}
\includegraphics[width=.5 \textwidth]{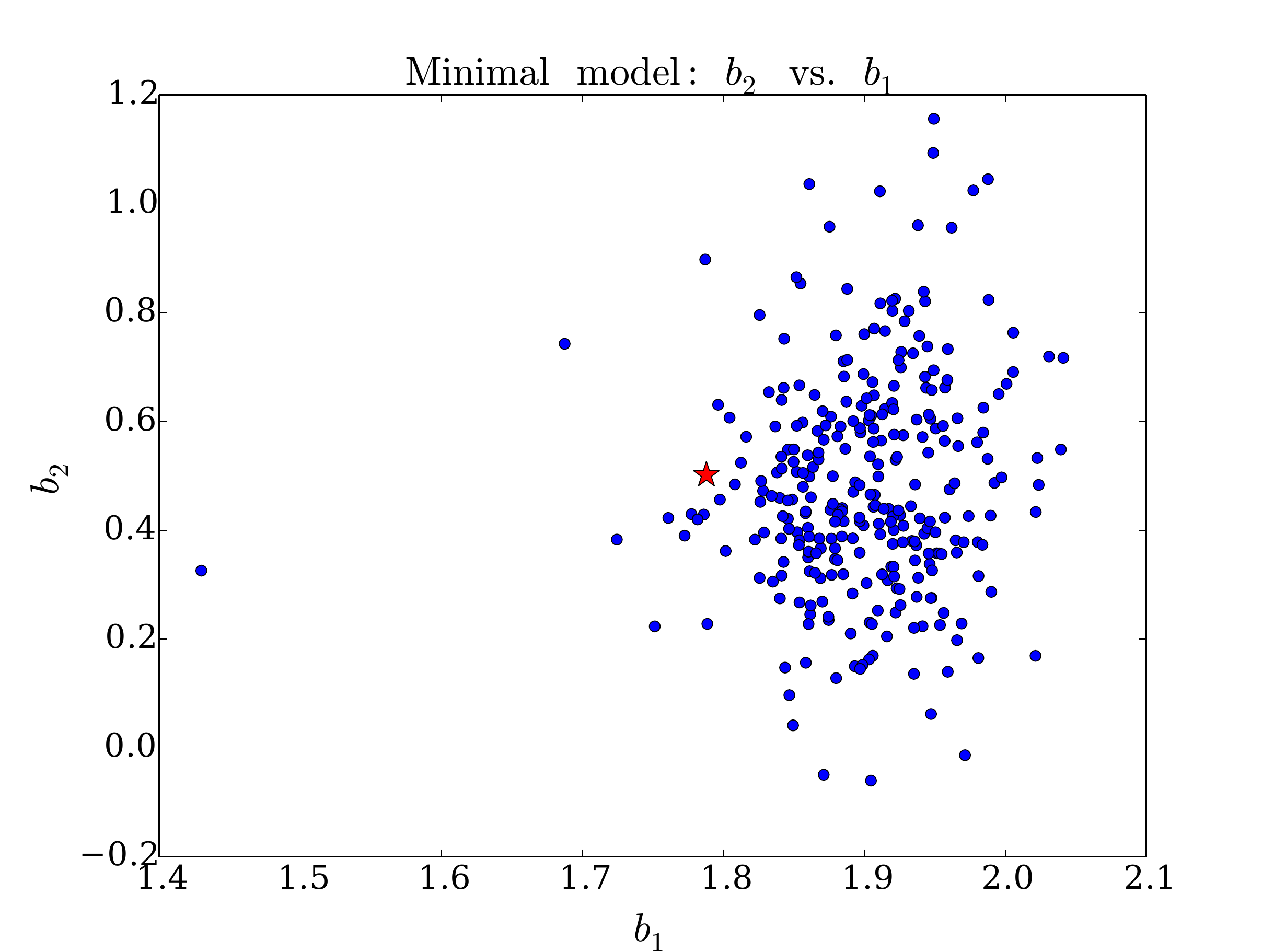}
\includegraphics[width=.465\textwidth]{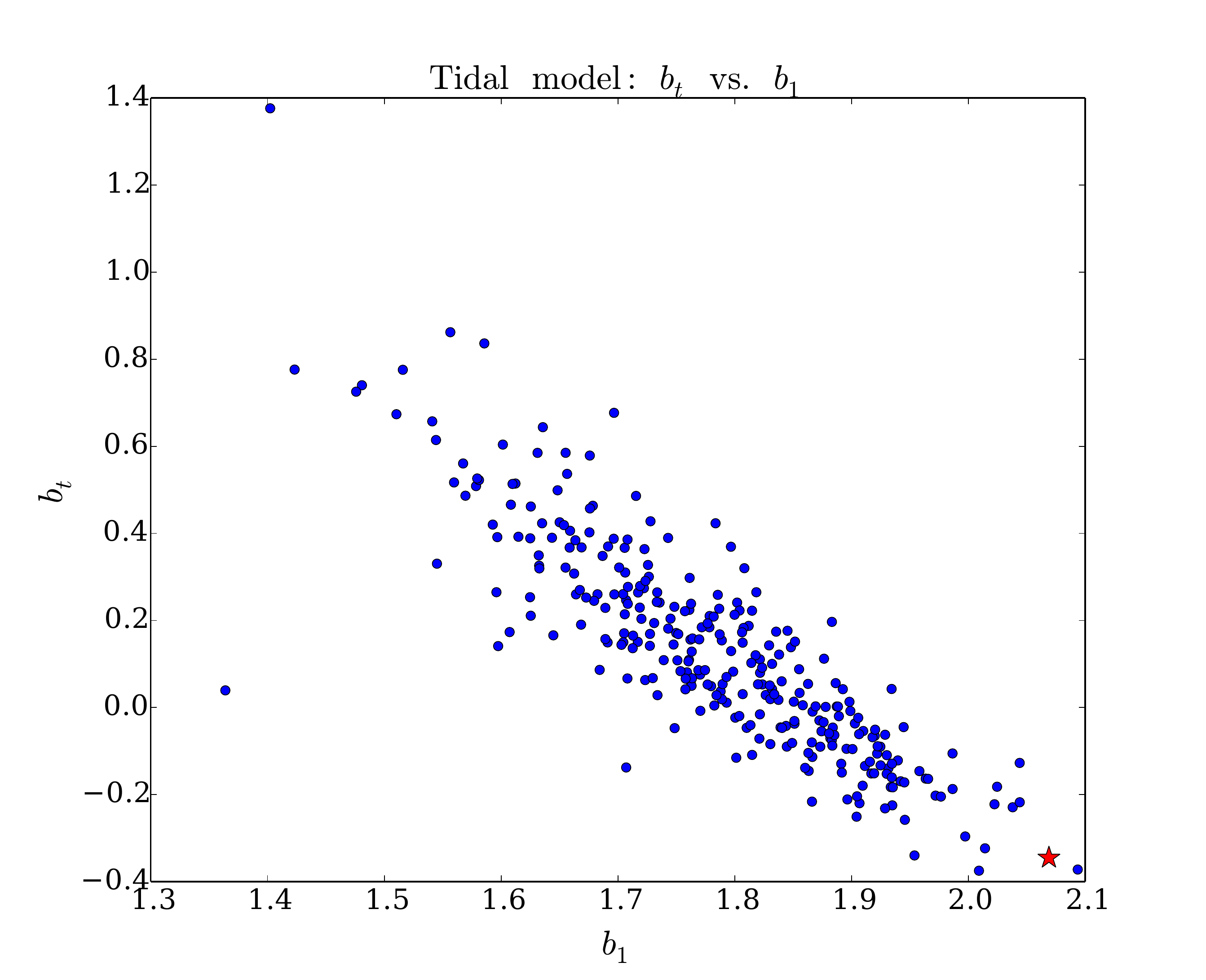}
\caption{The upper left panel shows the root mean square $\sigma(\alpha)$ versus $\left<\alpha \right>$ for the mocks. In all panels, the data values are marked with a red square. $\sigma(\alpha)$ is not highly correlated with $\left<\alpha \right>$. The upper right panel shows $b_1$ versus $\left<\alpha\right>$ for the mocks in the minimal model. Here $b_1$ is highly correlated with $\alpha$ as further discussed in \S\ref{sec:bias_parameters}. For the tidal tensor model $b_t$ decreases the degeneracy of $b_1$ and $\left<\alpha\right>$.  The lower left panel shows $b_2$ versus $b_1$ for the minimal model; again these biases are not highly correlated. The lower right panel shows $b_t$ versus $b_1$; $b_t$ is highly correlated with $b_1$, although in the mocks it does not follow the predicted $b_t=-(2/7)[b_1-1]$ relation; the slope is much steeper and the intercepts also disagree. Notably, the data value ($b_t=-0.35$) does come quite close to the $b_t$ predicted by this relation ($-0.31$).}
\label{fig:scatter_biases}
\end{figure*}

Our analysis held $\sigma_8$ fixed throughout; with this choice S15 and Gil-Mar\'in et al. (2015) both found  $2.6\%$ precision measurements of $b_1$. In the present work we find a $1\%$ precision measurement of $b_1$ in the minimal model and $4.0\%$ precision measurement of $b_1$ in the tidal tensor model.  It is not surprising that the constraint on $b_1$ is substantially worse in the tidal model: $b_1$ and $b_t$ are rather degenerate, as Figure \ref{fig:scatter_biases} shows (lower right panel). 

Our quoted error bars are consistent with the scatter of the linear bias values over the 298 mocks.  Note that to compute the error bar on the linear bias for each mock, we fix $\alpha$ to its mean value over all the mocks. Allowing $\alpha$ to float would introduce additional scatter into the measured $b_1$ values for the mocks since $\alpha$ and $b_1$ are highly degenerate, as the upper right panel of Figure \ref{fig:scatter_biases} shows. Since our goal is to use the scatter of the mocks' linear biases as a cross-check on our error bars for $b_1$ measured from the data, it is important to avoid introducing additional scatter from floating $\alpha$.

We do not obtain a strong constraint on the non-linear bias; the best-fit value is comparable to or smaller than the error bars both for the data and the mocks.  This poor constraint is because at lowest order in $\beta$, the non-linear bias only enters the $\ell=0$ pre-cyclic 3PCF multipole (see SE16b equation (21)), and this multipole is not as strong post-cyclically as $\ell=1$ and $\ell=2$ (see SE16b Figure 6).  While the non-linear bias does enter the pre-cyclic $\ell=2$ multipole (which is post-cyclically quite strong), it only does so at $\oO(\beta^2)$, and thus does not contribute greatly.  

Finally, we briefly note that our values of the integral constraint amplitude $c$ are consistent in sign with those found for the 3PCF of the same sample in the compressed basis of S15.  The magnitude is different at several sigma, however. Nonetheless, given that the integral constraint is also intended to marginalize over large-scale systematics, this difference in values is not highly surprising; the compressed basis of S15 and the full triangle bins used in the present work may respond differently to such systematics. The mocks show a scatter in $c$ of $0.009$, so the measurement of $c=-0.014$ in DR12 is plausible and does not
argue for observational systematics on very large scales.  A simple
calculation of the super-survey variance $\sigma_{\rm SS}^2$, based on the fluctuations of
mass in a sphere of volume equal to that of the survey, yields an
estimate of $\sigma_{\rm SS} = 0.004$.  This is of similar magnitude to the
scatter observed in the mocks.  The remaining factor of $2$ might be due
to mild degeneracies between $c$ and other parameters or simply to
this spherical estimate's being too optimistic.

\section{Conclusions}
\label{sec:concs}
In this work, we have used the novel 3PCF algorithm of SE15b to compute the 3PCF of $777,202$ LRGs from the CMASS sample of SDSS DR12.  This is the largest sample used for the 3PCF or bispectrum to date.  Using full triangles for a set of bins selected to avoid the regime where linear perturbation theory breaks down, we make the first high-significance detection of the BAO (respectively $4.5\sigma$ and $4.4\sigma$ for the two bias models we fit). Our previous work S15 measured the 3PCF of this sample in a compressed basis that integrated one triangle side over a wedge set by the other; while this approach had several physical motivations, it was likely sub-optimal for detecting BAO features. That previous work found a $2.8\sigma$ preference for the BAO.

With the present work's high significance BAO detection, we use the 3PCF to constrain the cosmic distance scale $\DV=2024\;{\rm Mpc}+29\;{\rm Mpc\;(stat)}+20\;{\rm Mpc\;(sys)}$ to redshift $z=0.57$ with $1.7\%$ precision (statistical plus systematic).  This distance measurement is the first use of the BAO method with the 3PCF.  We briefly explore the independence of the distance scale measured from the 3PCF relative to that measured from the 2PCF prior to reconstruction, and find they are essentially entirely independent. However, reconstruction is known to introduce some distance information from the 3PCF back into the 2PCF, and further work will explore the impact of reconstruction on the independence of the distance scales.  For the moment, we note that adding our distance scale to that measured from the unreconstructed 2PCF would produce a distance measurement with $1\%$ precision, comparable to the most recent precision of the distance scale measured from the reconstructed 2PCF (Cuesta et al. 2016) or power spectrum (Gil-Mar\'in et al. 2016).

Holding $\sigma_8$ fixed, we also place an extremely precise constraint on the linear bias, measuring $b_1$ to sub-percent accuracy in our minimal bias model.  This constraint is the most precise placed on the linear bias from either the 3PCF or the bispectrum, and is competitive with the most precise constraints on $b_1$ placed using any other techniques.  

Finally, we have made a moderate-confidence detection ($2.6\sigma$) of tidal tensor bias in agreement with the prediction of local Lagrangian biasing. Our error bars on the tidal tensor bias remain large because the tidal tensor bias is highly degenerate with the linear bias as they enter the 3PCF.  In future it will be worthwhile to explore avenues for breaking this degeneracy. The mocks do not appear to match the tidal tensor model well, but we do not expect this affects the BAO significance. In particular, the BAO significances are similar between our minimal and tidal tensor models, so the tidal tensor biasing is likely not a substantial driver of the BAO significance. We note that the bispectrum has been used before to test tidal tensor biasing on observational data; Feldman et al. (2001) found very mild evidence in favor of Eulerian biasing ($b_t = 0$) for the IRAS PSCz galaxy catalog. This galaxy population is very different from the LRGs considered in this work and so there is no conflict with the moderate-confidence detection of tidal tensor bias we report here.

In a companion paper (Slepian et al. 2016b), we use the same 3PCF dataset analyzed here with a different bias model to look for the imprint of baryon-dark matter relative velocities (Tseliakhovich \& Hirata 2010) in the late-time clustering of galaxies. This imprint can be an important possible systematic for BAO measurements from the 2PCF or power spectrum (Yoo, Dalal \& Seljak 2011), and as discussed in Slepian \& Eisenstein (2015a) has a unique signature in the 3PCF.  The companion paper finds that the 3PCF offers a $\sim 0.3\%$ constraint on any possible shift the relative velocity induces in the BAO scale inferred from the 2PCF, arguing for the 3PCF's utility in protecting future redshift surveys from this possible bias.

In sum, the same spectroscopic data sets currently used for 2PCF analyses with the BAO method can be used for 3PCF analyses with the BAO method. We hope the 3PCF will offer a new avenue to the cosmic distance scale. Used in conjunction with the 2PCF, we believe the 3PCF can increase the cosmological leverage of a given survey. With upcoming efforts such as DESI providing of order $30$ million spectra (Levi et al. 2013), taking full advantage of the BAO information in the 3PCF as well as that in the 2PCF will be highly desirable.


\section*{Acknowledgements}
We thank Blakesley Burkhart, Cora Dvorkin, Douglas Finkbeiner, Margaret Geller, Abraham Loeb, Philip Mocz, Ramesh Narayan, Stephen Portillo, Shun Saito, Marcel Schmittfull, Roman Scoccimarro,  Uro\v s Seljak, David Slepian, Ian Slepian, Joshua Suresh, Licia Verde, and Martin White for useful conservations. 

This material is based upon work supported by the National Science Foundation Graduate Research Fellowship under Grant No. DGE-1144152; DJE is supported by grant DE-SC0013718 from the U.S. Department of Energy. HGM acknowledges Labex ILP (reference ANR-10-LABX-63) part of the Idex SUPER, and received financial state aid managed by the Agence Nationale de la Recherche, as part of the programme Investissements d'avenir under the reference ANR-11-IDEX-0004-02. SH is supported by NSF AST1412966, NASA -EUCLID11-0004 and NSF AST1517593 for this work. WJP acknowledges support from the UK Science and Technology Facilities
Research Council through grants ST/M001709/1 and ST/N000668/1, the
European Research Council through grant 614030 Darksurvey, and the UK
Space Agency through grant ST/N00180X/1. GR acknowledges support from the National Research Foundation of Korea (NRF) through NRF-SGER 2014055950 funded by the Korean Ministry of Education, Science and Technology (MoEST), and from the faculty research fund of Sejong University in 2016. FSK thanks support from the Leibniz Society for the Karl-Schwarzschild fellowship. 

Funding for SDSS-III has been provided by the Alfred P. Sloan Foundation, the Participating Institutions, the National Science Foundation, and the U.S. Department of Energy Office of Science. The SDSS-III web site is http://www.sdss3.org/.

SDSS-III is managed by the Astrophysical Research Consortium for the Participating Institutions of the SDSS-III Collaboration including the University of Arizona, the Brazilian Participation Group, Brookhaven National Laboratory, Carnegie Mellon University, University of Florida, the French Participation Group, the German Participation Group, Harvard University, the Instituto de Astrofisica de Canarias, the Michigan State/Notre Dame/JINA Participation Group, Johns Hopkins University, Lawrence Berkeley National Laboratory, Max Planck Institute for Astrophysics, Max Planck Institute for Extraterrestrial Physics, New Mexico State University, New York University, Ohio State University, Pennsylvania State University, University of Portsmouth, Princeton University, the Spanish Participation Group, University of Tokyo, University of Utah, Vanderbilt University, University of Virginia, University of Washington, and Yale University.

\section*{References}
\hangindent=1.5em
\hangafter=1
\noindent Abazajian K. N. et al., 2009, ApJS, 182, 543.

\hangindent=1.5em
\hangafter=1
\noindent Aihara H. et al., 2011, ApJS, 193, 29.

\hangindent=1.5em 
\hangafter=1
\noindent Alam S. et al., 2016, preprint (arXiv:1607.03155).

\hangindent=1.5em 
\hangafter=1
\noindent Alam S. et al., 2015, ApJS, 219, 12

\hangindent=1.5em
\hangafter=1
\noindent Anderson et al. 2012, MNRAS 427, 4, 3435-3467.

\hangindent=1.5em
\hangafter=1
\noindent Anderson et al. 2014, MNRAS 441, 1, 24-62.

\hangindent=1.5em
\hangafter=1
\noindent Baldauf T., Seljak U., Desjacques V. \& McDonald P., 2012, PRD 86, 8.

\hangindent=1.5em 
\hangafter=1
\noindent Bernardeau F., Colombi S., Gazta\~{n}aga E., Scoccimarro R., 2002, Phys. Rep., 367, 1.

\hangindent=1.5em 
\hangafter=1
\noindent Blake C. \& Glazebrook K., 2003, ApJ 594, 2, 665-673.

\hangindent=1.5em
\hangafter=1
\noindent Blanton M. et al., 2003, AJ, 125, 2276.

\hangindent=1.5em 
\hangafter=1
\noindent Bolton, A. et al. 2012, AJ, 144, 144.

\noindent Bond J.R. \& Efstathiou G., 1984, ApJ 285, L45.

\noindent Bond J.R. \& Efstathiou G., 1987, MNRAS 226, 655-687.

\hangindent=1.5em 
\hangafter=1
\noindent Catelan P., Lucchin F., Matarrese S. \& Porciani C., 1998, MNRAS 297, 3, 692-712.

\hangindent=1.5em 
\hangafter=1
\noindent Catelan P., Porciani C. \& Kamionkowski M., 2000, MNRAS 318, 3, L39-L44.

\hangindent=1.5em
\hangafter=1
\noindent Chan K.C., Scoccimarro R. \& Sheth R.K., 2012, PRD 85, 8, 083509.

\hangindent=1.5em
\hangafter=1
\noindent Cuesta A.J. et al., 2016, MNRAS 457, 2, 1770-1785. 

\hangindent=1.5em 
\hangafter=1
\noindent Dawson, K. et al. 2013, AJ, 145, 10

\hangindent=1.5em
\hangafter=1
\noindent  Doi M. et al., 2010, AJ, 139, 1628.

\hangindent=1.5em 
\hangafter=1
\noindent Eisenstein, D.J. et al. 2011, AJ, 142, 72

\hangindent=1.5em
\hangafter=1
\noindent Eisenstein D.J. \& Hu W., 1998, ApJ 496, 605.

\hangindent=1.5em
\hangafter=1
\noindent Eisenstein D.J., Hu W. \& Tegmark M., 1998, ApJ  504:L57-L60.

\hangindent=1.5em
\hangafter=1
\noindent Eisenstein D.J., Seo H.-J., Sirko E. \& Spergel D., 2007, ApJ 664, 2, 675-679.

\hangindent=1.5em
\hangafter=1
\noindent Eisenstein D.J., Seo H.-J. \& White M., 2007, ApJ 664, 2, 660-674.

\hangindent=1.5em
\hangafter=1
\noindent Feldman H., Frieman J., Fry J.N. \& Scoccimarro R., 2001, PRL 86, 8, 1434-1437.

\hangindent=1.5em
\hangafter=1
\noindent Fry J.N., 1996, ApJL 461, L65.

\hangindent=1.5em
\hangafter=1
\noindent Fukugita M. et al., 1996, AJ, 111, 1748.

\hangindent=1.5em
\hangafter=1
\noindent  Gazta\~naga E., Cabr\'e A., Castander F., Crocce M. \& Fosalba P., 2009, MNRAS 399, 2, 801-811.

\hangindent=1.5em
\hangafter=1
\noindent Gil-Mar\'in H. et al., 2016, MNRAS, doi:10.1093/mnras/stw1264.

\hangindent=1.5em
\hangafter=1
\noindent Gil-Mar\'in H., Nore\~na J., Verde L., Percival W.J., Wagner C., Manera M. \& Schneider D.P., 2015, MNRAS 451, 1, 539-580.

\hangindent=1.5em
\hangafter=1
\noindent Groth E. J. \& Peebles P. J. E., 1977, 217, 385.

\hangindent=1.5em
\hangafter=1
\noindent Gunn J.E. et al., 1998, AJ, 116, 3040.

\hangindent=1.5em 
\hangafter=1
\noindent Gunn, J.E. et al. 2006, AJ, 131, 2332

\hangindent=1.5em 
\hangafter=1
\noindent Hamilton A.J.S, 1998, in ``The Evolving Universe: Selected Topics on Large-Scale Structure and on the Properties of Galaxies,'' Dordrecht: Kluwer.

\hangindent=1.5em
\hangafter=1
\noindent Holtzmann J.A., 1989, ApJS 71, 1.

\hangindent=1.5em 
\hangafter=1
\noindent Hu W. \& Haiman Z., 2003, PRD 68, 6, 063004.

\noindent Hu W. \& Sugiyama N., 1996, ApJ 471:542-570.

\hangindent=1.5em
\hangafter=1
\noindent Kitaura F.-S. et al., 2016, PRL 116, 7, 171301.

\hangindent=1.5em
\hangafter=1
\noindent Kitaura F.-S. \& He{\ss} S., 2013, MNRAS 435, 1, L78-L82.

\hangindent=1.5em
\hangafter=1
\noindent Kitaura F.-S., Yepes G. \& Prada F., 2014, MNRAS 439, L21.

\hangindent=1.5em
\hangafter=1
\noindent Kitaura F.-S., Gil-Mar\'in H., Sc\'occola C.G., Chuang C.-H., M\"{u}ller V., Yepes G. \& Prada F., 2015a, MNRAS 450, 2, 1836-1845.

\hangindent=1.5em
\hangafter=1
\noindent Kitaura F.-S. et al., 2015b, preprint (arXiv:1509.06400).

\hangindent=1.5em
\hangafter=1
\noindent Lewis A., 2000, ApJ, 538, 473.

\hangindent=1.5em 
\hangafter=1
\noindent Linder E.V., 2003, PRD 68, 8, 083504.

\hangindent=1.5em
\hangafter=1
\noindent Lupton R., Gunn J.E., Ivezi\'c Z., Knapp G. \& Kent S., 2001, ``Astronomical Data Analysis Software and Systems X'', v. 238, 269.

\hangindent=1.5em
\hangafter=1
\noindent McDonald P., PRD 74, 10, 103512.

\hangindent=1.5em
\hangafter=1
\noindent McDonald P. \& Roy A., 2009, JCAP 0908, 020.

\hangindent=1.5em
\hangafter=1
\noindent Mehrem R., 2002, preprint (arXiv:0909.0494v4). 

\hangindent=1.5em 
\hangafter=1
\noindent Olver W. J., Lozier D. W., Boisvert R. F., Clark C. W. eds. NIST Handbook
of Mathematical Functions. Cambridge University Press, Cambridge,\\
Available at http://dlmf.nist.gov/

\hangindent=1.5em
\hangafter=1
\noindent Padmanabhan N. et al., 2008, ApJ, 674, 1217.

\noindent Peebles P.J.E. \& Yu J.T., 1970, ApJ 162, 815.

\hangindent=1.5em
\hangafter=1
\noindent Percival W.J. et al., 2014, MNRAS 439, 3, 2531-2541.

\hangindent=1.5em
\hangafter=1
\noindent Pier J.R. et al., 2003, AJ, 125, 1559.

\hangindent=1.5em
\hangafter=1
\noindent Planck Collaboration, Paper XIII, 2015, preprint (arXiv:1502.01589).

\hangindent=1.5em
\hangafter=1
\noindent Planck Collaboration, Paper XVII, 2015, preprint (arXiv:1502.01592).

\hangindent=1.5em
\hangafter=1
\noindent Reid B. et al., 2016, MNRAS 455, 2, 1553-1573.

\hangindent=1.5em
\hangafter=1
\noindent Rodr\'iguez-Torres S. et al., 2015, preprint (arXiv:1509.06404).

\hangindent=1.5em
\hangafter=1
\noindent Ross A.J. et al., 2012, MNRAS 424, 1, 564-590.

\hangindent=1.5em
\hangafter=1
\noindent Ross A.J. et al., 2013, MNRAS 428, 2, 1116-1127.

\hangindent=1.5em
\hangafter=1
\noindent Ross A.J. et al., 2014, MNRAS 437, 2, 1109-1126.

\hangindent=1.5em
\hangafter=1
\noindent Ross A.J. et al., 2015, in prep.

\hangindent=1.5em
\hangafter=1
\noindent Sakharov A.D., 1966, Soviet Journal of Experimental and Theoretical Physics 22, 241.

\hangindent=1.5em
\hangafter=1
\noindent Schmittfull M.M., Feng Y., Beutler F., Sherwin B. \& Yat Chu, M., 2015, preprint (arXiv:1508.06972).

\hangindent=1.5em
\hangafter=1
\noindent Scoccimarro R., Couchman H.M.P. \& Frieman J.A., 1999, ApJ 517:531-540.

\noindent Seo H.J. \& Eisenstein D.J., 2003. ApJ 598, 2, 720-740.

\hangindent=1em
\hangafter=1
\noindent Slepian Z. \& Eisenstein D.J., 2015a, MNRAS 448, 1, 9-26.

\hangindent=1em
\hangafter=1
\noindent Slepian Z. \& Eisenstein D.J., 2015b, MNRAS 454, 4, 4142-4158.

\hangindent=1em
\hangafter=1
\noindent Slepian Z. \& Eisenstein D.J., 2015c, MNRASL 455, 1, L31-L35.

\hangindent=1em
\hangafter=1
\noindent Slepian Z. \& Eisenstein D.J., 2016a, MNRAS 457, 24-37. 

\hangindent=1em
\hangafter=1
\noindent Slepian Z. \& Eisenstein D.J., 2016b, preprint (arXiv:1607.03109). 

\hangindent=1em
\hangafter=1
\noindent Slepian Z.  et al., 2015, preprint (arXiv:1512.02231). 

\hangindent=1em
\hangafter=1
\noindent Slepian Z.  et al., 2016b, preprint (arXiv:--). 

\hangindent=1em 
\hangafter=1
\noindent Smee, S.A. et al. 2013, AJ, 126, 32

\hangindent=1em
\hangafter=1
\noindent Smith J.A. et al., 2002, AJ, 123, 2121.

\noindent Sunyaev R.A. \& Zel'dovich Ya. B., 1970, Ap\&SS 7, 3.

\hangindent=1.5em
\hangafter=1
\noindent Szapudi I., 2004, ApJ, 605, L89.

\hangindent=1.5em
\hangafter=1
\noindent Tseliakhovich D. \& Hirata C. 2010, PRD, 82, 083520.

\hangindent=1.5em
\hangafter=1
\noindent Xu X., Padmanabhan N., Eisenstein D.J., Mehta K.T. \& Cuesta A.J., 2012, MNRAS 427, 3, 2146-2167.

\hangindent=1.5em
\hangafter=1
\noindent Yoo J., Dalal N. \& Seljak U., 2011, JCAP, 7, 018.

\hangindent=1.5em
\hangafter=1
\noindent York D.G. et al., 2000, AJ, 120, 1579.

\end{document}